%% file: main.tex
\documentclass[a4paper,11pt]{article}
\input{preamble}
\title{\textbf{\LARGE{Co-Design of Aeroelastic Systems with Deep Reinforcement Learning}}}

\author[1]{Yao Cheng Li$^*$}
\author[1]{Urban Fasel}

\affil[1]{\small Department of Aeronautics, Imperial College London, SW7 2AZ, United Kingdom}
\affil[$*$]{{\footnotesize corresponding author: paul.li20@imperial.ac.uk}}

\begin{document}
\date{}
\maketitle
\vspace{-.5in}

\begin{abstract}
Control co-design considers the physical system and its controller together, enabling the strong coupling between system design and control to be uncovered and exploited. This is especially relevant in aeroelastic flight systems, where structural, aerodynamic, and control design choices jointly determine manoeuvrability and efficiency. This paper presents a model-free nested co-design framework for aeroelastic systems using deep reinforcement learning, in which a design-conditioned control policy is trained with proximal policy optimisation while an outer loop updates a distribution over candidate design parameters. The approach is evaluated on three case studies of increasing complexity: a spring-mass-damper system, a pitch-plunge-flap aerofoil, and a highly flexible high-aspect-ratio glider performing a thermal-soaring mission in a stochastic environment. Across these case studies, the framework is shown to progressively concentrate the design search towards high-performing regions and to outperform policies trained on randomly sampled designs. The results also show that reward shaping plays an important role in enabling stable learning in partially observed and stochastic environments. In the final glider case, the method jointly addresses wing design, flight control, and mission-level behaviour in the presence of aeroelastic coupling and atmospheric uncertainty. These results highlight the potential of model-free co-design for complex aeroelastic systems in which design, control, and mission objectives are tightly coupled.
\end{abstract}

\section{Introduction}
\label{sec: Introduction}
Co-Design, or control co-design, is a multidisciplinary optimisation approach that uses nested or simultaneous optimisation loops to consider both the physical system and its controller from the initial stages of design, enabling solutions that may not be achievable when the two are optimised separately \cite{Garcia-Sanz_CCD_Game_Changer, Fathy_Plant_controller_coupling, Herber_Nested_simultaneous_CCD}. In this paper, we present a co-design framework for thermal-soaring fixed-wing gliders that jointly optimises airframe design and control policy in the presence of aeroelastic coupling and environmental uncertainty. The central premise is that, for highly flexible high-aspect-ratio gliders, structural design choices and control decisions are strongly coupled, together determining manoeuvrability and efficiency. To address this, we develop a low-fidelity but fully parameterisable aeroelastic glider simulation and combine it with deep reinforcement learning to enable integrated system-controller co-design optimisation. 

From complex multidisciplinary engineering systems such as aircraft \cite{Jaddivada_Codesign_Electric_Aircraft} and autonomous robots \cite{Zardini_Codesign_Autonomous_Systems} to societal-scale engineered infrastructures such as electrical grids \cite{Liu_Codesign_Electrical_Grids}, co-design has been used to uncover and exploit the inherent interdependence between system architecture and control. As highlighted in the strategic future of control \cite{Alleyne_Control_Roadmap_2030}, there is an increasing need for a unifying integrated framework for system architecture and control optimisation, particularly with the adoption of emerging approaches such as data-driven control. This need becomes especially pronounced in systems with strong dynamic coupling, where system parameters directly affect controllability and observability, thereby constraining or enabling different control strategies \cite{Silva_Applications_of_CCD_Survey}. Aeroelastic systems are a representative example of such systems.

Aeroelasticity is the study of the coupling between aerodynamic, structural elastic, and inertial forces acting on aerodynamic bodies such as wings, tails and fuselages \cite{Collar_Triangle}. As the global aviation industry is projected to double its emissions by 2050 if no action is taken \cite{Bergero_Pathways_to_NetZero_Aviation, Almena_Reducing_environmental_impact_of_aviation}, there has been significant research and development into technologies that can reduce overall emissions. One active area is the development of high aspect ratio wings, such as NASA's transonic truss-braced wing \cite{Hariton_TTBW} and various Airbus high-aspect-ratio wing prototypes \cite{Wilson_Airbus_SAH}. It is known from Prandtl's lifting line theory \cite{Hunsaker_Prandtl_lifting_line} and past aircraft designs \cite{Kretov_Evaluation_of_efficiency_of_HARW} that increasing aspect ratio reduces induced drag; however, it also introduces important challenges. High-aspect-ratio wings can incur significant structural mass penalties to support long cantilever wings, reduce lateral controllability through increased roll damping, and increase susceptibility to aeroelastic flutter and divergence. Although some of these effects can be mitigated through alternative design choices, such as span loading strategies demonstrated on X-HALE \cite{Cesnik_X-HALE}, such choices may introduce secondary consequences. For example, they can increase roll inertia and strengthen coupling between roll, pitch and elastic degrees of freedom when elevon-based control is used.

These competing design objectives become even more pronounced in unpowered soaring flight, where aerodynamic efficiency and manoeuvrability are both critical. Soaring flight is conducted by birds \cite{Richardson_Albatross_flight, Harel_Vultures_thermal_soaring}, gliders \cite{Pennycuick_Crosscountry_soaring} and more recently UAVs \cite{Cowling_UAVs_thermal_soaring} to maintain or even gain altitude without propulsion. Different types of soaring flight exploit different atmospheric phenomena to extract energy: dynamic soaring exploits wind-shear gradients in turning climb and dive manoeuvres \cite{Zhao_Optimal_glider_dynamic_soaring}, gust soaring exploits separated boundary layers caused by terrain features using similar manoeuvres \cite{Richardson_Albatross_flight}, and thermal soaring exploits rising thermals generated by local atmospheric convection \cite{Pennycuick_Crosscountry_soaring}. Manoeuvrability is placed at a premium for all types of soaring: high load factors are required at the bottom of the dive, minimum-radius turns are needed to remain within narrow thermal columns, and traversing in zigzag patterns is needed above jagged ridge lines where thermals arise from. At the same time, the glide ratio of an unpowered aircraft in still air is determined by its lift-to-drag ratio. Hence, a balance must be struck between aerodynamic efficiency and manoeuvrability.

Thermal soaring introduces an additional challenge of locating thermals and planning trajectories under uncertainty. Whilst the mechanism of thermal formation is well understood, accurately predicting the precise location of a thermal in practice remains difficult. Cumulus clouds can indicate thermals because they form as a by-product of condensing water vapour brought to altitude by a thermal \cite{Cowling_UAVs_thermal_soaring}. Thermals may also be produced above local hot-spots such as fires or hot surfaces with high thermal emissivity, including concrete and rocks \cite{Cowling_UAVs_thermal_soaring}. However, such visual cues are not always reliable because thermals are highly dynamic. In practice, the only reliable method of detecting a thermal is to fly through or near it, where it can be detected using a variometer or inferred by the pilot from changes in stick force caused by the local flow field. Consequently, thermal soaring requires a continual balance between exploration and exploitation in a highly uncertain and dynamic environment. Prolonged exploration for thermals may cause long detours and altitude loss; however, successfully finding thermals provides the opportunity to convert gained altitude into kinetic energy and complete the planned flight path sooner.

Deep reinforcement learning, a subfield of machine learning, is an emerging approach for decision-making and control under uncertainty while balancing exploration and exploitation. Agents learn through repeated trial-and-error interactions with an environment, enabling them to explore different trajectories before converging towards behaviours that maximise long-term reward \cite{Li_DRL_overivew}. Deep reinforcement learning has been applied in many complex domains, including robotic control \cite{Mahmood_RL_robots}, video games \cite{Silver_Alphago, Vinyals_Alphastar, Mnih_Atari_DRL}, protein structure prediction \cite{Jumper_Alphafold2, Abramson_Alphafold3}, and sustainable aviation rollout prediction \cite{Muir_Sustainable_aviation_rollout}. More recently, it has also seen application in co-design \cite{Sadat_Two_Timescale_RL_CCD, Schaff_N-limb, Schaff_Codesign} and soaring-related problems \cite{Flato_Principles_of_autonomous_thermal_soaring, Cui_Soaring_strategy_RL, Notter_DRL_updraft_mapping_exploitation}. However, the intersection of aeroelastic glider design, thermal-soaring decision-making, and integrated co-design remains unexplored.

These considerations motivate the development of a co-design framework for a fixed-wing glider operating in a thermal soaring environment using deep reinforcement learning. Specifically, in this study, we 
\begin{enumerate}
    \item create a low-fidelity flight simulation with a fully parameterisable elastic high aspect ratio fixed-wing glider model;
    \item develop control strategies using deep reinforcement learning for highly flexible aircraft; and
    \item integrate these elements into a unified data-driven co-design framework for the joint optimisation of glider design and control.
\end{enumerate}

In the remainder of the paper, we first discuss related work in Section \ref{sec: Related Works}, before presenting the proposed nested co-design approach in Section \ref{sec: Method}. We then evaluate the method on three case studies of increasing complexity in Section \ref{sec: results and discussion}, before concluding in Section \ref{sec: conclusion and outlook}.

%----------------------------------------------------------------------------------------

\section{Related Works}
\label{sec: Related Works}

Although multidisciplinary design optimisation (MDO) in aircraft design shares many features with co-design, control design has often been treated separately, particularly in problems where time-domain dynamic interactions are not central to the design task. By contrast, co-design has seen extensive and successful application in fields such as robotics \cite{Silva_Applications_of_CCD_Survey}, where platform morphology and system parameters strongly influence the associated control strategy. This is especially true when deep reinforcement learning (RL) is used to address problems that are too complex to solve with classical optimisation methods. For example, Schaff et al. developed a nested co-design optimisation framework to jointly optimise robotic design and control policy on benchmark walking robot problems \cite{Schaff_Codesign}. In their approach, the model-free stochastic algorithm proximal policy optimisation (PPO) was used to learn a control policy within the inner loop, while a Gaussian mixture model (GMM) parameterised the design distribution in the outer loop. They showed that the learned walking gait of several robotic morphologies outperformed baseline designs. This framework was later extended to arbitrary morphologies using discrete design parameters \cite{Schaff_N-limb}. Similarly, Sadat et al., following the control-inspired co-design paradigm proposed in \cite{Garcia-Sanz_CCD_Game_Changer}, presented a two-timescale nested co-design optimisation framework on the benchmark EcoRacer problem \cite{Sadat_Two_Timescale_RL_CCD}. In that work, the RL algorithm Q-learning with temporal difference was used in both the inner control loop and the outer design loop, but with different hyperparameters to separate the learning rates across the two timescales. Their results demonstrated convergence towards the system optimal solution. Other nested co-design approaches have also been explored, including evolutionary optimisation of generic creature morphology with PPO-based inner-loop control \cite{Gupta_Unimal}, graph heuristic search in the outer loop with graph neural network control in the inner loop \cite{Zhao_Robogrammar}, and differentiable formulations in which design parameters are embedded directly within the control policy and jointly optimised \cite{Chen_Hardware_as_policy}.

Within the domain of aerial vehicles, \cite{Bergonti_CCD_winged_drones} demonstrated a co-design framework for a morphing winged drone navigating an obstacle course. In that work, the NSGA-II evolutionary algorithm was used to evolve a population pool of candidate designs over many generations, while a trajectory optimiser minimised energy consumption and mission time subject to constraints such as hardware limits, obstacle avoidance, and initial flight conditions. Likewise, \cite{Zhao_Co-design_aerial_robot_graph_grammar} developed a joint UAV and controller parameter optimisation framework using a hierarchical optimisation algorithm, with a heuristic graph search in the outer loop over mixed continuous and discrete UAV designs, and gradient-based sequential quadratic programming in the inner loop for LQR controller tuning. Both studies demonstrated improved performance over fixed baseline aircraft designs. Part of this improvement can be attributed to the use of morphing mechanisms in flexible aircraft, which have been shown to significantly enhance manoeuvring characteristics \cite{Ajanic_Sharp_turns_with_morphing, Wuest_Agile_perching}. Related ideas have also appeared in flexible structural systems. For example, \cite{Maraniello_Codesign_flexible_structures} applied a single-shooting method to the co-design of an actuated flexible pendulum, jointly optimising structural geometry and a control policy parametrised using discrete sine series and spline basis functions. Their results showed improved objective function values compared with optimising the control policy alone. Taken together, these studies show the value of integrated design and control optimisation for flexible and morphing systems, but comparable co-design studies for aeroelastic gliders in soaring flight remain limited.

Although there is a lack of co-design in soaring applications, several studies have demonstrated the use of RL algorithms for thermal searching and tracking. For example, \cite{Flato_Principles_of_autonomous_thermal_soaring} used deep deterministic policy gradient (DDPG) to train an agent that can track a thermal column under windy conditions. Because stable flight and thermal tracking are difficult to learn simultaneously, reward shaping was introduced to address learning bottlenecks by providing intermediate rewards for desirable behaviours in addition to the thermal tracking mission-level reward. Similarly, \cite{Cui_Soaring_strategy_RL} used the soft-actor critic algorithm to explore a stochastic thermal field, showing that the agent learned distinct exploration and exploitation behaviours. Unlike these earlier studies, \cite{Notter_DRL_updraft_mapping_exploitation} introduced a recurrent network layer after the fully connected neural network layers and demonstrated an improved ability to encode thermal updraft dynamics during flight. However, these soaring studies were all implemented using simplified rigid flight dynamics models, in which variables such as angle of attack and roll angle are determined directly by the control policy rather than emerging from higher-fidelity vehicle dynamics. As a result, important flight characteristics such as short-period pitching oscillations and Dutch roll behaviour are not represented. These characteristics are directly relevant to co-design, particularly when the objective is to produce aircraft configurations that are dynamically stable and trimmable. More broadly, RL algorithms such as PPO may also learn slowly and become stuck in a local optimum in more complex problems, depending on policy initialisation and hyperparameter selection. For example, \cite{Coletti_PPO_warm_start} addressed this issue through imitation learning to warm-start policy training for rigid aircraft control with full control dynamics.

A key component of any co-design framework is a well-defined dynamical system, which serves as the environment in which the agent is trained. In \cite{Flato_Principles_of_autonomous_thermal_soaring, Cui_Soaring_strategy_RL, Notter_DRL_updraft_mapping_exploitation}, the equations of flight mechanics were implemented directly as the training environment. More generally, dedicated flight simulation environments have also been developed, such as FlightMare, which combines modular Gazebo-based simulation with Unity engine visualisation for computational efficiency \cite{Song_Flightmare}, as well as modified general-purpose environments such as PyFlyt, a UAV simulation environment built on Pybullet \cite{Tai_Pyflyt}. As highlighted in \cite{Kaup_RL_physics_engine_review, Erez_RL_env_comparison}, the suitability of a training environment depends not only on efficiency, accuracy, and physics capabilities, but also on usability, accessibility and extensibility for the broader research community. Across these criteria, both studies identified Mujoco (multi-joint dynamics with contact) as one of the strongest available frameworks. Although Mujoco has traditionally been used for multi-jointed robotic systems with contact dynamics, as in \cite{Gupta_Unimal, Schaff_Codesign}, it has also been applied more broadly, from simulating fruit fly flight and locomotion \cite{Vaxenburg_Fruit_fly} to modelling an unborn fetus using soft body dynamics \cite{Kim_Human_fetus}. For the present work, implementing a flight dynamics environment in Mujoco provides not only an efficient and accurate physics-based RL training environment, but also a wider accessible basis for future extensions of aeroelastic flight co-design within a wider simulation ecosystem.

%----------------------------------------------------------------------------------------

\section{Method}
\label{sec: Method}

\begin{figure}[!t]
	\centering
	\includegraphics[width=\textwidth]{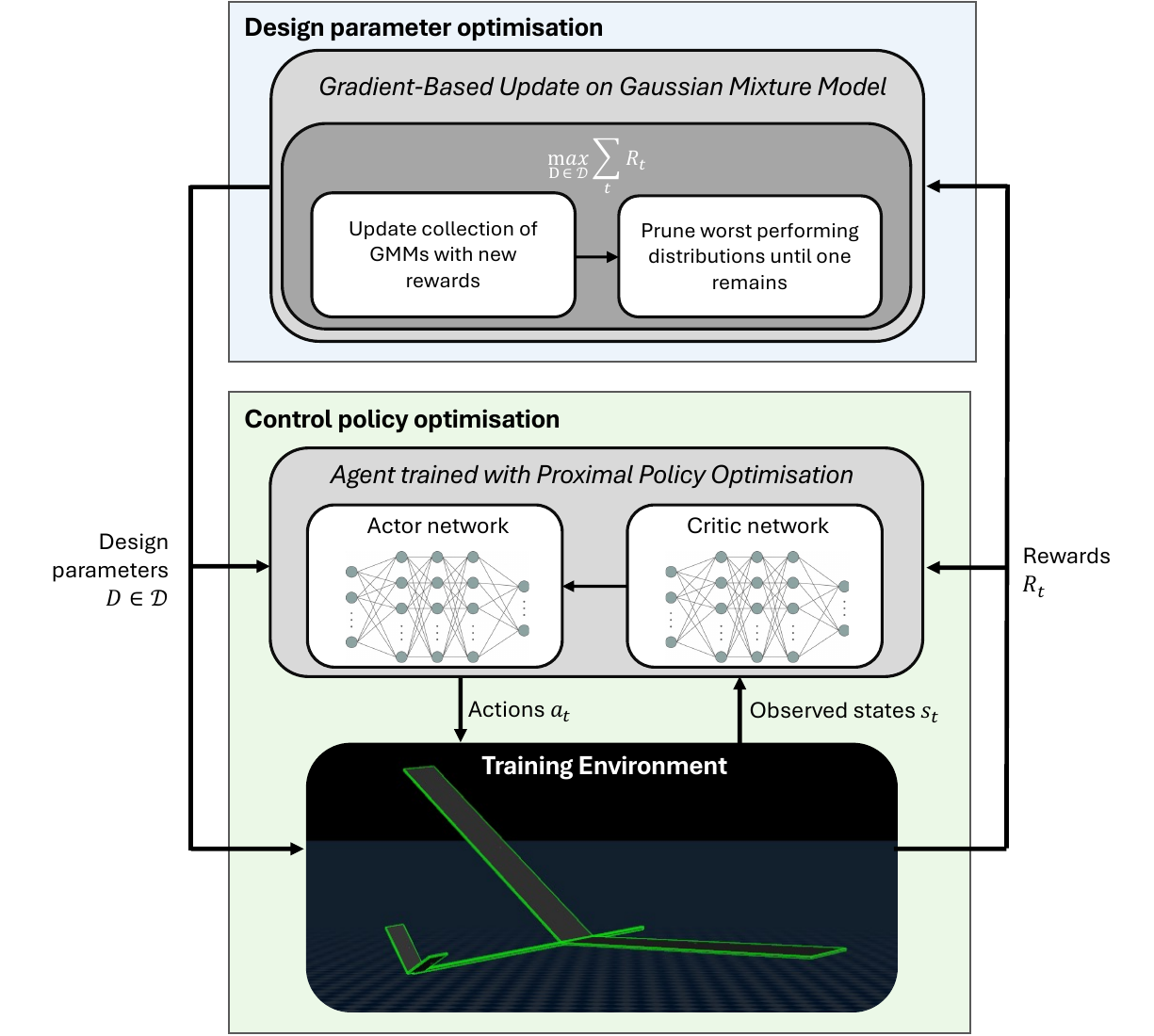}
    %\vspace{-15pt}
    \caption{Nested co-design framework for joint glider and controller optimisation. A design-conditioned control policy is trained in the inner loop using reinforcement learning, while the outer loop updates and prunes a distribution over candidate designs according to performance. This progressively concentrates training on high-performing regions of the design space.}
	\label{fig: Co-design flow}
\end{figure}

To apply co-design to a flexible fixed-wing glider in a thermal soaring mission, this study adopts the model-free co-design framework from Schaff et al. \cite{Schaff_Codesign} illustrated in Figure \ref{fig: Co-design flow}. The method maintains a distribution over candidate design parameters and alternates between sampling designs from this distribution and improving a control policy through interaction with the environment. Crucially, the control policy is conditioned on the sampled design parameters, such that a single policy is trained across the full range of candidate designs rather than learning a separate controller for each design instance.

This formulation can be viewed as a nested optimisation loop, in which the control policy optimisation is performed in the inner loop and the design parameter optimisation in the outer loop. For each sampled design, the inner-loop performs $N_{inner}$ training iterations before handing off to the outer design parameter optimisation loop. The inner loop is not trained to convergence at every outer loop iteration, as both loops must be updated jointly. Doing so also reduces the total computational cost. During this process, the inner loop receives design parameters $D$ from the outer loop and returns the achieved reward information through interaction with the environment. The outer loop optimiser seeks to maximise the total reward $\sum_t R_t$ by updating the design parameters within the feasible design space $\mathcal{D}$. Meanwhile, the inner loop optimiser updates the control policy to maximize the discounted reward $\sum^\infty_{t=0} \gamma^t R_t$ where $\gamma \in (0, 1)$ is the discount factor that weights short-term rewards more heavily than those received further in the future.

The inner loop control problem is formulated as a design-conditioned Markov decision process (MDP). At a given time $t$, the environment is in a state $s_t \in \mathcal{S}$, the agent selects an action $a_t \in \mathcal{A}$, and the system transitions to the next state $s_{t+1}$ while producing a scalar reward $r_t$. Unlike a conventional MDP, the transition function is also conditioned on the design parameter provided by the outer-loop optimiser. The one step transition probability of a given trajectory $\tau$, which defines a transition from $s_t$ to $s_{t+1}$ with action $a_t$, is therefore defined as

\begin{equation}
    P(s_{t+1}, s_t, a_t, D) = Pr(s'=s_{t+1} | s=s_t, a=a_t, d=D),
\end{equation}

where $D$ remains fixed throughout a single episode and is only updated when the episode is reset. The rest of the formulation follows the standard RL setting, except that both the transition dynamics and the control policy are conditioned on the design parameters. The expected return is therefore written as

\begin{equation}
    \underset{\tau \sim \pi}{\mathbb{E}} [R(\tau)] = \int_{\tau} P(\tau|\pi)R(\tau),
\end{equation}

where the compounded transition probability is defined as

\begin{equation}
    P(\tau|\pi) = \rho_0 (s_0) \prod^{T-1}_{t=0} P(s_{t+1} | s_t, a_t, D) \pi(a_t | s_t, D),
\end{equation}

and $\rho_0(s_0)$ is the probability distribution of starting at a given initial state $s_0$.

For the inner-loop policy optimisation, we employ the on-policy, policy-gradient-based optimisation algorithm proximal policy optimisation (PPO) \cite{Schulman_PPO}, which seeks to maximise a clipped surrogate advantage function defined as

\begin{equation}
    \label{eqn: PPO loss}
    \hat{\mathbb{E}}_t \left[ \min(r_t(\theta)\hat{A}_t, \text{clip}(r_t(\theta), 1 - \epsilon, 1 + \epsilon)\hat{A}_t) \right],
\end{equation}

where $r_t(\theta) = \frac{\pi_\theta(a|s,D)}{\pi_{\theta_k}(a|s,D)}$ is the relative difference between the updated policy and the old policy $\pi_{\theta_k}$. The advantage function $\hat{A}_t = Q^\pi(s_t,a_t,D) - V^\pi(s_t,D)$ quantifies the advantage of sampled actions $a_t \sim \pi_{\theta}(\cdot|s_t,D)$, collected from $N_\text{step} \times N_\text{env}$ number of trajectories, by taking the difference between the state-action value function and the state value function. The clipped surrogate robustifies the algorithm against large policy updates and thereby improves training stability.

\begin{algorithm}
\footnotesize
\caption{Nested co-design optimisation algorithm.} \label{algo: Co-design}
    \begin{algorithmic}[0]
        \Require\parbox[t]{0.8\linewidth}{Initial policy network $\pi_\theta$, initial $N_\text{design}$ multivariate Gaussian distributions, \\ and initial state observations $s_0$ from $N_\text{env}$ environments.}
        \State $\text{step} \gets 0$
        \State Allocate $N_\text{design}$ distributions to $N_\text{env}$ environments.
        \While{$\text{step} < N_\text{total steps}$}
            \For{$i \gets 1$ to $N_\text{prune}$}
                \State For each environment, sample design parameters $D_n \sim \mathcal{N}(\mu_n, \Sigma_n)$ from its assigned distribution.
                \For{$j \gets 1$ to $N_\text{inner}$}
                    \For{$k \gets 1$ to $N_{step}$}
                        \State Sample an action $a_t \sim \pi_\theta(\cdot|s_t,D_n)$.
                        \State Step through environment and collect batches of trajectories $\langle s_t, a_t, r_t, s_{t+1} \rangle$.
                        \State $\text{step} \gets \text{step}+N_\text{env}$.
                    \EndFor
                    \State Update policy $\pi_\theta$ with PPO to minimise Eq. \ref{eqn: PPO loss} using collected rollout trajectories.
                \EndFor
                \If{$\text{step} > N_\text{freeze}N_\text{step}N_\text{env}$}
                    \State Update design distributions with Adam to minimise Eq. \ref{eqn: Schaff loss}.
                \EndIf
            \EndFor
            \If{$\text{step} > N_\text{freeze}N_\text{step}N_\text{env}$ \textbf{and} $N_\text{design} > 1$}
                \State Evaluate all distributions by sampling $N_\text{eval}$ designs and 
                \State selecting deterministic actions $a_t = \underset{a}{\text{argmax}} \space \pi_\theta(a|s_t,D_n)$ during evaluation.
                \State Prune the worst performing distributions based on the prune rate $\alpha_\text{prune}.$
                \State Resample and re-allocate $N_\text{design}$ distributions to $N_\text{env}$ environments.
            \EndIf
        \EndWhile
    \end{algorithmic}
\end{algorithm}

To further stabilise training, the outer-loop optimiser should progressively narrow the design search space of $D$ to limit the total transitions $P(\tau|\pi)$ that the policy must learn. This motivates the use of Gaussian mixture models (GMMs), which represent the design distribution using $N_{\text{design}}$ clusters. These clusters are progressively pruned over time based on the evaluated reward and prune rate $\alpha_\text{prune}$ \cite{Schaff_Codesign}, allowing the inner-loop training to focus increasingly on high-performing design parameters. Each $n$ Gaussian model parameterises the design space through a mean vector $\mu_n$ and covariance matrix $\Sigma_n$, with the mean constrained such that $\mu \in \mathcal{D}$. During training, $N_{\text{env}}$ environments are instantiated, each allocated a design parameter $D_n$ sampled from the corresponding GMM distribution, with density

\begin{equation}
    f_{\mu_n, L_n}(D) = \mathcal{N}(\mu_n, \Sigma_n).
\end{equation}

To satisfy the symmetric positive definiteness requirement of the covariance matrix, we reparameterised the covariance matrix using the Cholesky decomposition such that $\Sigma_n=L_nL_n^T$, with diagonal elements constrained to $L_{ii} > 0$. After $N_\text{inner}$ rollout updates, we update the parameters of the GMMs using the Adam optimiser \cite{kingma2014adam}, a gradient-based optimiser for stochastic functions, to minimise the weighted negative log-likelihood loss

\begin{equation}
    \label{eqn: Schaff loss}
    \mathcal{L}(\mu_n, L_n) = -\frac{1}{N} \sum^N_{i=1} \left( \sum_t R_t \right)_i \log f_{\mu_n, L_n}(D_i),
\end{equation}

where $N$ is the total number of episodes taken between distribution updates.

After every $N_\text{prune}$ distribution updates, the current design distributions are evaluated using newly sampled designs, and the worst-performing distributions are pruned based on the mean reward achieved. This focuses the inner loop training on an increasingly smaller range of design parameters and promotes more stable convergence. However, even with a robust policy update algorithm such as PPO, policy training may still fail without additional stabilisation \cite{Coletti_PPO_warm_start}. To address this, a warm start training step was introduced that freezes the outer-loop optimiser for $N_{\text{freeze}}$ initial inner-loop updates before initiating the co-design process. This prevents poor-performing policies during the early iterations of the inner loop from corrupting the optimisation in the outer loop. Furthermore, the means of the initial $N_{\text{design}}$ distributions are sampled using Latin Hypercube sampling (LHS) to ensure uniform initial coverage of the design space $\mathcal{D}$. Each covariance matrix is initialised as a diagonal matrix of variances set to 10\% span of the parameter range. The complete algorithm is outlined in Algorithm \ref{algo: Co-design}.

%----------------------------------------------------------------------------------------

\section{Results \& Discussion}
\label{sec: results and discussion}

We apply the co-design framework introduced in Section \ref{sec: Method} to three case studies of increasing complexity: a forced spring-mass-damper system, a pitch-plunge-flap aerofoil system, and finally a fixed-wing glider executing a thermal soaring mission.

\subsection{Spring-Mass-Damper System}
\label{sec: sdc}

To demonstrate the co-design workflow, we first apply it to a simple spring-mass damper system governed by

\begin{equation}
    m\ddot{x}(t) + c\dot{x}(t) + kx(t) = F(t),
\end{equation}

where the mass $m$, damping coefficient $c$ and spring stiffness $k$ parameterise the system. An external force $F(t)$ is applied such that

\begin{align}
    F(t) &= F_\text{max} a_n, \quad \text{for } t\in[n \Delta t, (n+1) \Delta t).
\end{align}

The goal is to jointly optimise a control policy and system parameters $D = \{ k, c \}$, subjected to $k, c \in [-1, 1]$, to minimise the position tracking error. The policy is sampled at constant time intervals $\Delta t$ to obtain the desired action $a_n \sim \pi_\theta(\cdot|s_n, D)$ with $a_n \in [-1,1]$. We generate a series of random step reference signals with varying step size $x_{\text{ref},i}$ that are separated by random time intervals $\Delta \tau_i$ such that

\begin{align}
    x_\text{ref} &= x_{\text{ref}, i}, \quad \text{for } t \in [\tau_i, \tau_{i+1}), \\
    \tau_i &= \tau_{i-1} + \Delta \tau_i.
\end{align}

We initialise $\tau_0 = 0$ and sample from a uniform distribution such that

\begin{align}
    x_{\text{ref},i} &\sim \mathcal{U}(-x_{\text{ref}_{\max}}, x_{\text{ref}_{\max}}), \\
    \Delta \tau_i &\sim \mathcal{U} (\Delta \tau_{\min}, \Delta \tau_{\max}).
\end{align}

The reference signal is chosen to provide sufficient variance during training, allowing the agent to learn from a wide range of trajectories, such an example signal can be seen in Figure \ref{fig: ppf simulation}. The environment reward at each step is defined as $R_n = -(x_\text{ref} - x)^2$.

\begin{figure}[!tb]
	\centering
	\includegraphics[width=0.6\textwidth]{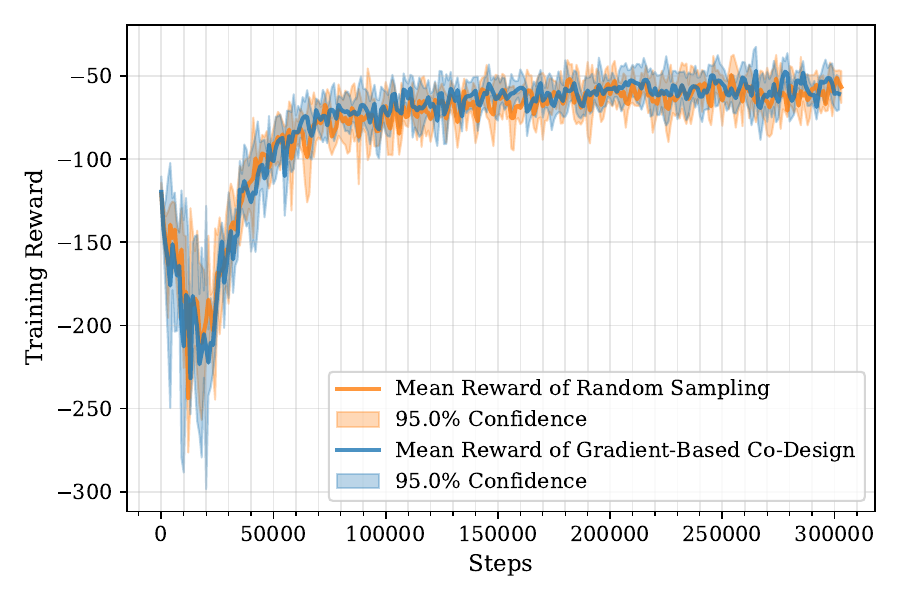}
	%\vspace{10pt}
	\caption{Smoothed reward curve of the mass-spring-damper system repeated over 5 training attempts.}
    \vspace{30pt}
	\label{fig: sdc reward}
\end{figure}

\begin{figure}[!tb]
    \centering
    \begin{minipage}{0.49\textwidth}
        \centering
        \includegraphics[width=1\textwidth]{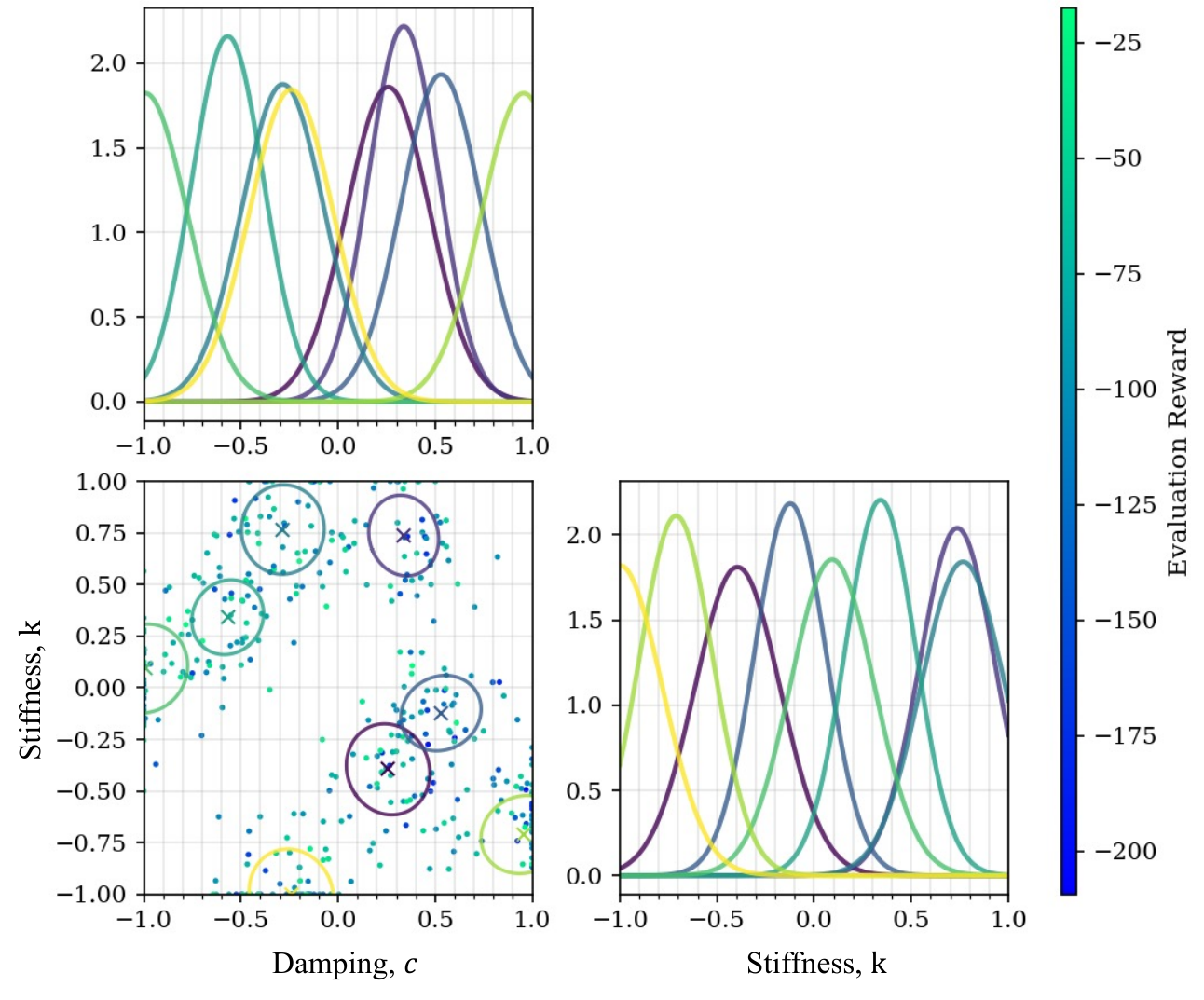}
        %\vspace{10pt}
        \captionsetup{justification=centering}
        \caption{Co-design evaluation stage for pruning at training step 61440.}
        \label{fig: SDC GMM 1}
    \end{minipage}\hfill
    \begin{minipage}{0.49\textwidth}
        \centering
        \includegraphics[width=1\textwidth]{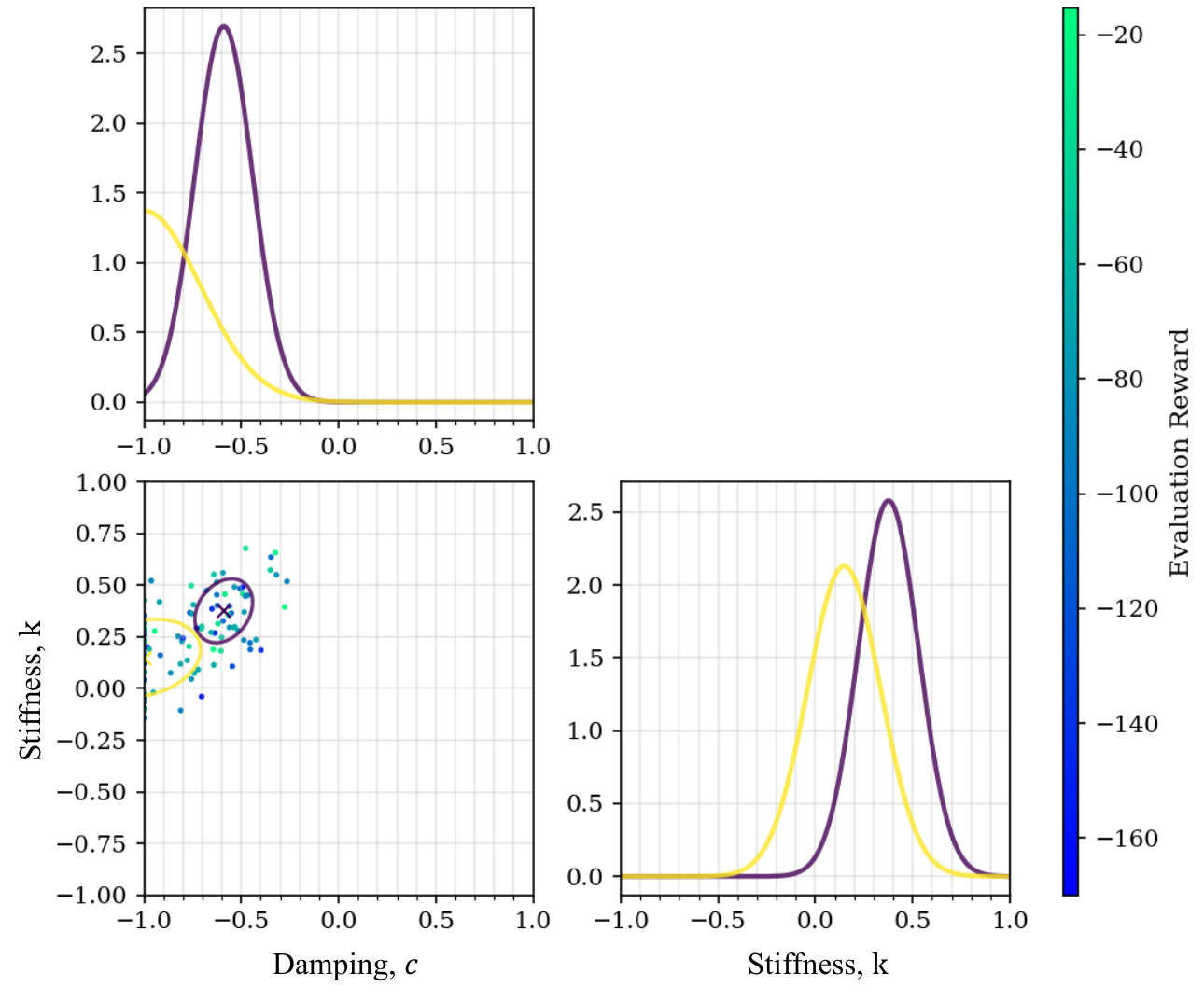}
        %\vspace{10pt}
        \captionsetup{justification=centering}
        \caption{Co-design evaluation stage for pruning at training step 184320.}
        \label{fig: SDC GMM 2}
    \end{minipage}
\end{figure}

For this application, we train on $N_\text{env}=16$ parallelised environments for a total of $N_\text{total steps}=300000$ steps. The optimisation is initialised with $N_\text{design} = 8$ distributions, an evaluation and prune interval of $N_\text{prune} = 5$, inner-loop update interval of $N_\text{inner} = 3$, a policy update interval $N_\text{step} = 256$, and a prune rate $\alpha_\text{prune} = 0.5$. The policy is pre-trained for $N_\text{freeze} = 10$ inner-loop update before initiating the co-design process. The training process is repeated over five independent runs, and the aggregated reward curve is shown in Figure \ref{fig: sdc reward}. On a simple problem such as a mass-spring-damper system, the performance of unguided random sampling is comparable to the guided learning co-design algorithm, resulting in little difference between the reward curves.

The resulting optimisation behaviour is illustrated in Figures \ref{fig: SDC GMM 1} and \ref{fig: SDC GMM 2}, which show the Gaussian mixture models together with the rewards obtained by designs sampled from their distributions for one representative training run. During the early stage of training, design parameters $D$ are sampled from a large number of distributions, enabling broad exploration of the design domain $\mathcal{D}$, as shown in Figure \ref{fig: SDC GMM 1}. At this stage, each distribution is represented by only a small number of environments, so the inner-loop optimiser must learn across a broad range of transition dynamics from limited samples per design region. As training progresses, the number of distributions is reduced to 1, indicating that the optimisation has concentrated on a candidate design region. Figure \ref{fig: SDC GMM 2} shows the final evaluation before the last pruning step, where the remaining two distributions are concentrated around the region of negative damping and positive spring stiffness. Within this actuator-constrained case study, negative damping accelerates the transition between consecutive step signals, while positive stiffness helps prevent the system from straying too far from the bounded region of reference signals $x_{\text{ref}_{\max}} = 2$. Overall, this example shows that the proposed model-free co-design framework can progressively narrow the design distribution and identify interpretable parameter regions associated with improved reward during training.

\subsection{Pitch-Plunge-Flap Aerofoil}
\label{sec: ppf}

\begin{figure}[!tb]
	\centering
	\includegraphics[width=0.5\textwidth]{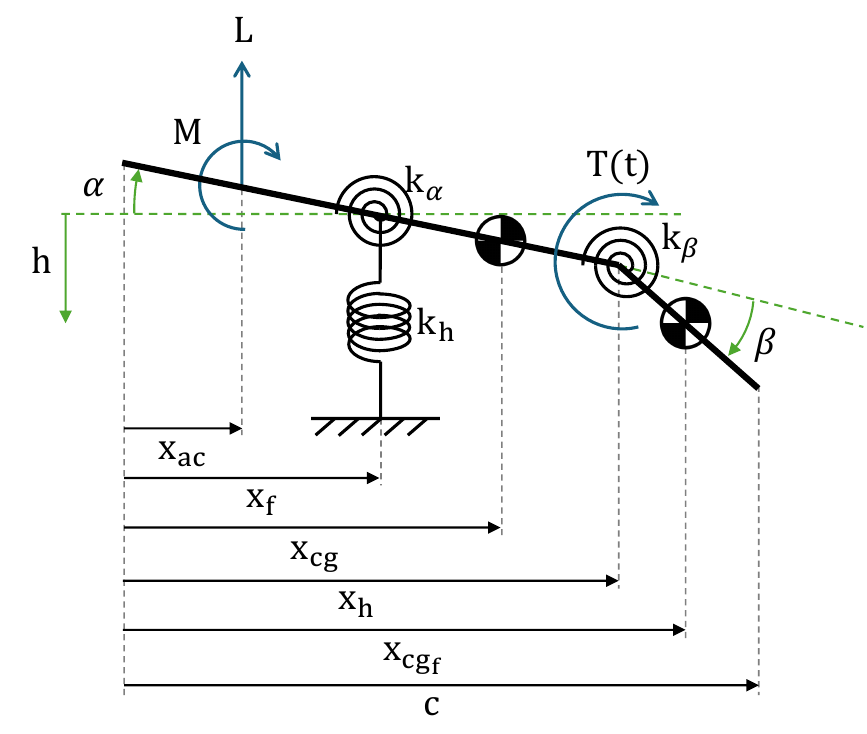}
	%\vspace{10pt}
	\caption{Schematic of the pitch-plunge-flap aerofoil.}
    \vspace{10pt}
	\label{fig: ppf diagram}
\end{figure}

Next, we apply the co-design framework to an idealised thin aerofoil with plunge, pitch and flap degrees of freedom, shown in Figure \ref{fig: ppf diagram}. These three degrees of freedom are analogous to the out-of-plane bending, torsional twist, and flap degrees of freedom on a 3D cantilever wing. This second case study is intended to assess whether the framework remains effective in a partially observed aeroelastic system with critical stability boundaries inside the design space. Although the pitch-plunge-flap aerofoil state-space model is relatively straightforward to analyse using conventional methods, it introduces several important challenges for model-free learning approaches. First, the hidden aerodynamic states are not included in the observation, making the control problem partially observed. Second, the aerofoil undergoes several critical bifurcations within the chosen design space and operating condition. Specifically, it exhibits aeroelastic phenomena including flutter, divergence, and control reversal, substantially increasing the complexity of the optimisation problem.

Here, we adopt a quasi-steady aerodynamic formulation based on \cite{theodorsen1949general, dimitriadis2017introduction}, which introduces six additional hidden aerodynamic states and makes the problem partially observed, as defined by

\begin{align}
    (\mathbf{A}_\text{ae} + \rho \mathbf{B}_\text{ae}) \ddot{\mathbf{y}}
    + (\mathbf{C}_\text{ae} + \rho U \mathbf{D}_\text{ae}) \dot{\mathbf{y}}
    + (\mathbf{E}_\text{ae} + \rho U^2 \mathbf{F}_\text{ae}) \mathbf{y} + \rho U^3 \mathbf{Ww}
    &= \begin{bmatrix}
        0 & 0 & T(t)
    \end{bmatrix}^T, \\
    \dot{\mathbf{w}} - \mathbf{W}_1 \mathbf{y} - U \mathbf{W_2} \mathbf{w} &= 0,
\end{align}

where $\mathbf{y} = \begin{bmatrix} h & \alpha & \beta \end{bmatrix}^T$ and $\mathbf{w} = \begin{bmatrix} w_1 & \dots & w_6 \end{bmatrix}^T$ define the degrees of freedom and hidden aerodynamic states, respectively. We refer the reader to \cite{dimitriadis2017introduction} for the derivation of the aeroelastic matrices. Unlike the implementation in \cite{dimitriadis2017introduction}, we formulate the inertial properties as lumped masses with parameterisable location, defined by non-dimensional positions $\bar{x}_{cg_w}$ and $\bar{x}_{cg_f}$ for the main wing section and trailing edge flap. Coordinates are non-dimensionalised by the semi-chord such that $\bar{x} = (x - c/2)/(c/2)$. Furthermore, a motor torque $T(t)$ is applied on the flap hinge such that

\begin{equation}
    T(t) = T_\text{max} a_n, \quad \text{for } t\in[n \Delta t, (n+1) \Delta t).
\end{equation}

The design variables in this co-design problem are the main wing centre of gravity $\bar{x}_{cg_w}$, the pitch spring stiffness $k_\alpha$, and the flap hinge stiffness $k_\beta$. These parameters directly affect the aeroelastic stability boundaries of the system, including the onset of flutter, divergence, and control reversal, thereby motivating their joint optimisation with the control policy. The objective to be minimised is the tracking error $h_e(t) = h_\text{ref}(t) - h(t)$ where the reference signal $h_\text{ref}(t)$ is prescribed in the same way as in Section \ref{sec: sdc}. Furthermore, during training, the freestream velocity is sampled from $U \sim \mathcal{U}(10, 80)$. Over the prescribed design space $\mathcal{D}$ and velocity range $U$, the system may exhibit aeroelastic instabilities for certain parameter combinations, resulting in an unstable system that naturally oscillates. The system is also torque-limited, with $T_{\max}=5$, such that even though a high flap hinge stiffness $k_\beta$ can delay the flutter onset velocity, it can cause the controller to saturate and prevent accurate tracking of the reference point. To account for this effect, in addition to the tracking error, the reward function $R_n$ also penalises excessive oscillations, such that

\begin{equation}
    R_n = - \left| h_\text{ref} - h \right| - \lambda_{\dot{\beta}} \dot{\beta}^2 + 0.5,
\end{equation}

where $\lambda_{\dot{\beta}} = 0.01$ is the penalty weight on the flap deflection rate. Without this term, the optimiser would minimise the tracking error $h_e(t)$ alone, without accounting for oscillatory behaviour associated with flutter. This highlights the importance of incorporating domain knowledge into the reward structure. In addition, an episode is terminated early if the plunge state $h$ exceeds twice the maximum expected reference value $h_{\text{ref}_{\max}}=1$, or if either angular states exceed $\alpha > \pi/4$ or $\beta > \pi/4$. Combined with the constant $0.5$ reward, this termination criterion guides the optimisation towards stable designs and helps prune candidates that deviate too far from the expected operating range. The co-design hyperparameters used in the remaining case studies are presented in Appendix \ref{sec: codesign hyperparameters}.

\begin{figure}[!tb]
	\centering
	\includegraphics[width=0.7\textwidth]{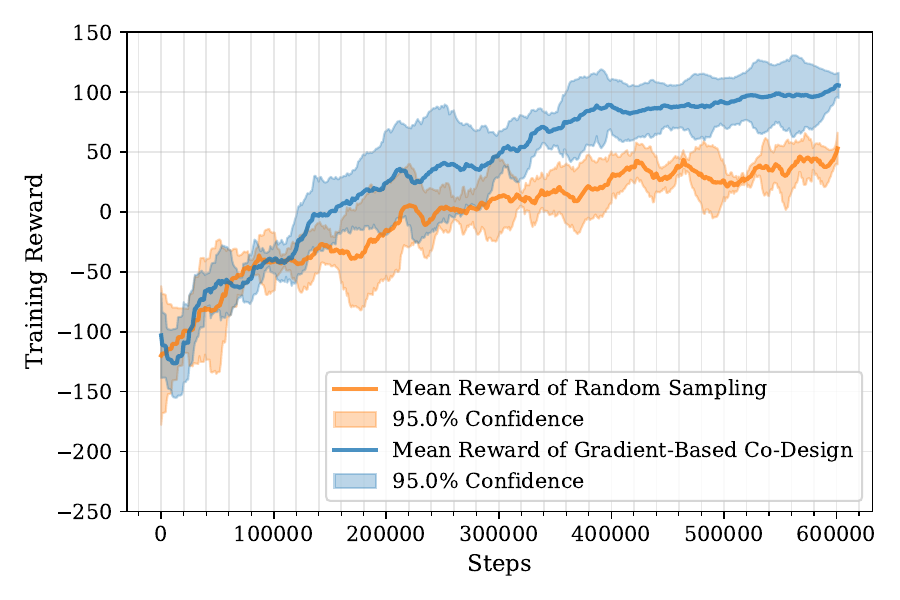}
	%\vspace{10pt}
	\caption{Smoothed reward curve of the pitch-plunge-flap system repeated over 5 training attempts.}
    %\vspace{10pt}
	\label{fig: ppf reward}
\end{figure}

\begin{figure}[!tb]
	\centering
	\includegraphics[width=0.8\textwidth]{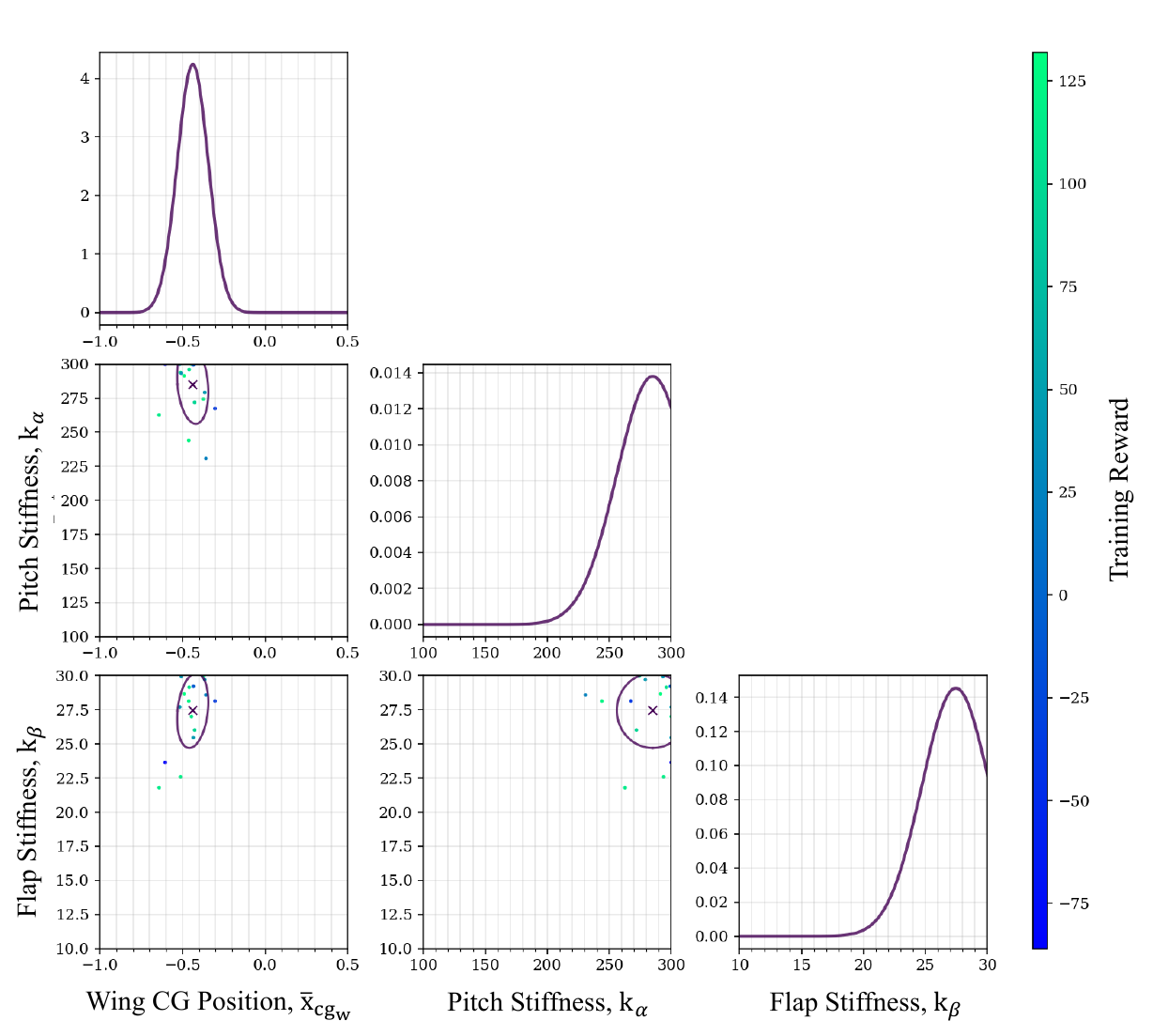}
	%\vspace{10pt}
	\caption{The training reward landscape before the final design distribution update of the pitch-plunge-flap system.}
    \vspace{20pt}
	\label{fig: ppf gmm}
\end{figure}

\begin{figure}[!tb]
	\centering
	\includegraphics[width=0.7\textwidth]{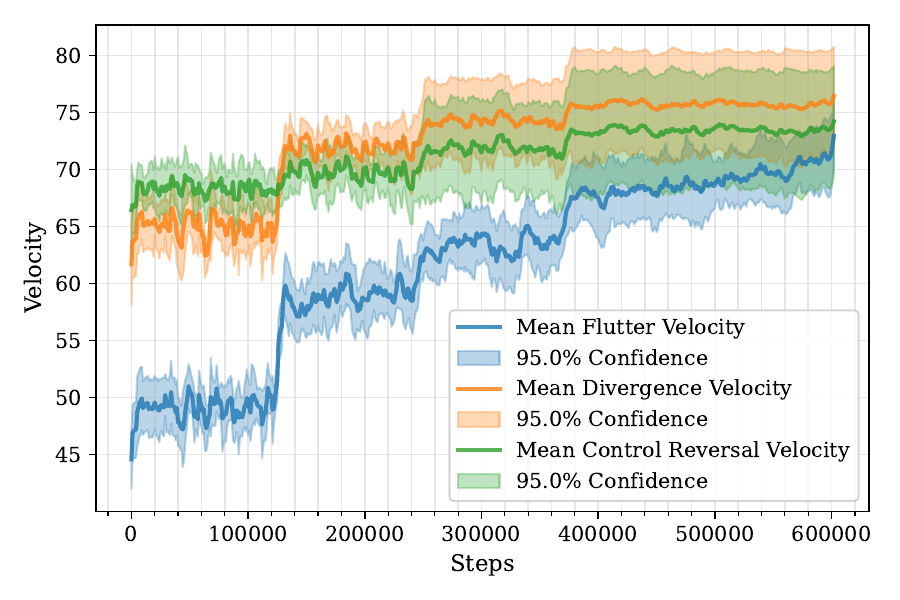}
	%\vspace{10pt}
	\caption{Evolution of flutter, divergence and control reversal speeds over the iterations of the co-design algorithm.}
    \vspace{20pt}
	\label{fig: ppf critical velocities}
\end{figure}

\begin{figure}[!tb]
	\centering
	\includegraphics[width=0.6\textwidth]{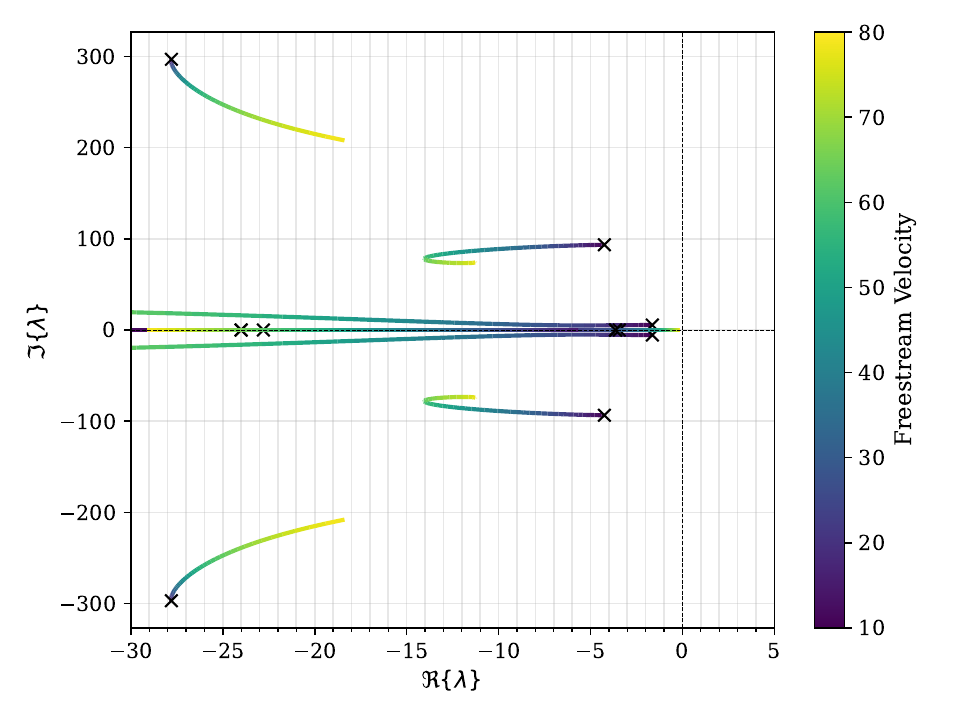}
	%\vspace{10pt}
	\caption{Root locus plot of the optimised aeroelastic system across velocity range $U \in (10, 80)$.}
    \vspace{20pt}
	\label{fig: ppf root locus}
\end{figure}

\begin{figure}[!tb]
	\centering
	\includegraphics[width=0.8\textwidth]{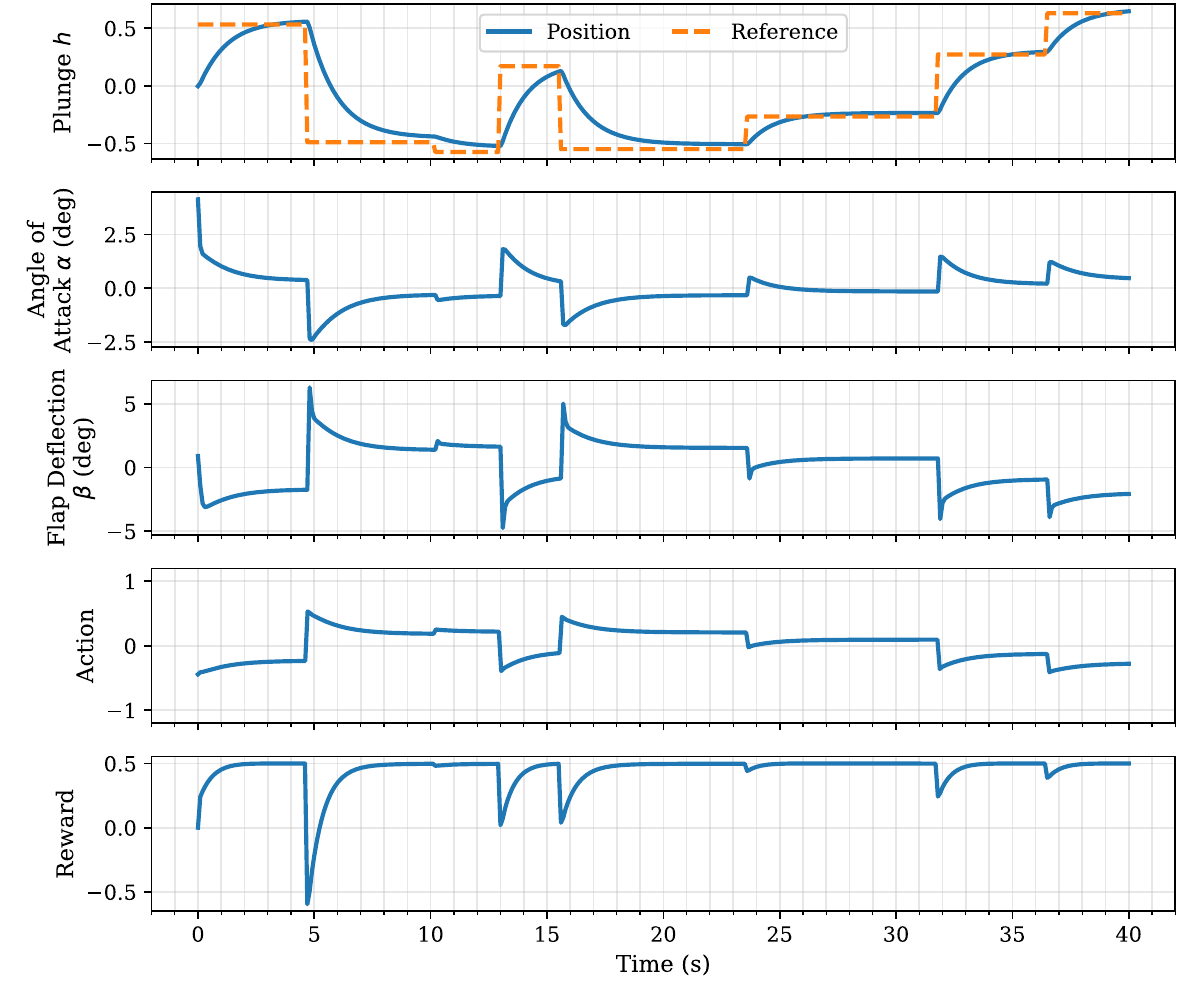}
	%\vspace{10pt}
	\caption{Evaluating a trained policy on an example generated reference signal.}
    \vspace{10pt}
	\label{fig: ppf simulation}
\end{figure}

Figure \ref{fig: ppf reward} shows the training reward curve across five independent instances, together with a benchmark PPO policy trained on randomly sampled designs from $\mathcal{D}$ using LHS. Whilst random sampling exposes the policy to the full design space, it also requires the policy to learn over a much broader range of transition dynamics, rather than focusing on a small subset of higher-performing designs. As a result, the random-sampling baseline achieves substantially lower reward and converges more slowly than the guided co-design algorithm. Figure \ref{fig: ppf gmm} shows the final remaining Gaussian distribution from one representative training instance, with the mean centred around the design parameters $\bar{x}_{cg_w} = -0.54$, $k_\alpha = 254.16$, and $k_{\beta} = 30.00$. This design pushes both spring stiffness parameters closer to their upper limits and shifts the main-wing centre of gravity ahead of its flexural axis at $\bar{x}_f = -0.2$. Together, these design choices delay the onset of the three critical velocities, as shown in Figure \ref{fig: ppf critical velocities}. The best sampled design also remains stable within the specified operational velocities as shown in the Figure \ref{fig: ppf root locus}. However, the system does still exhibit control reversal at 78.9 m/s, although this critical velocity has also been shifted close to the upper bound of the operating regime.

Overall, the co-design algorithm converges towards a region of the design space where the system remains stable and avoids any critical system bifurcation behaviour. With suitable reward shaping, a model-free co-design approach can identify effective design-control combinations with relatively limited explicit constraints on the optimisation process or additional domain-specific knowledge. However, depending on the complexity of the problem, the agent may struggle to learn all possible system dynamics, even after progressively reducing the design space. Under certain conditions, the learned policy may exhibit unintuitive behaviours, such as the non-zero steady state errors as shown in Figure \ref{fig: ppf simulation}. Unlike model-based or optimal control solutions, these behaviours are not always easy to interpret. Model-free RL approaches may therefore be most appropriate for problems that are too difficult to solve using conventional methods, or for problems where broad design exploration is the primary objective.

\subsection{Thermal Soaring of a Glider}
\label{sec: co-aegis}

Finally, we apply the co-design framework to a flexible fixed-wing aircraft model implemented in Mujoco, as detailed in Appendix \ref{sec: Mujoco Flight Environment}. In this case study, the design is based on the Feather RC glider from Aerosandbox \cite{Sharpe_Aerosandbox}, and we introduce the wing area $S_\text{ref} \in [0.04, 0.3]$ and wing aspect ratio $AR \in [4, 24]$ as design variables such that $D = \{S_\text{ref}, AR\}$. The baseline glider from \cite{Sharpe_Aerosandbox} is optimised solely for minimal energy loss during glide, and has an $S_\text{ref}=0.240$ and $AR=17.25$. The objective is to jointly optimise these parameters and the control policy for a glider performing a thermal soaring mission, where success is measured by the number of randomly prescribed waypoints reached within a limited mission time. 

The chosen design space induces substantial variation in the aircraft dynamics and handling characteristics, including large variations in roll damping and best glide speeds. Optimising the wing area and aspect ratio introduces several coupled effects. For example, increasing the wing area increases wing mass and inertia, as shown in equations \ref{eqn: wing mass} and \ref{eqn: wing inertia}, while also increasing the total lift force generated through equation \ref{eqn: wing stability forces} and increasing the wing stiffness through equations \ref{eqn: wing area moment of inertia} and \ref{eqn: wing polar moment of inertia}. This in turn would affect the handling qualities of the aircraft, such as shifting the flight envelope to a lower speed due to its lower wing loading, increasing overall aircraft inertia and mass, and increasing the aircraft's susceptibility to gusts.

Because the aircraft is unpowered, it must learn not only to follow the waypoint sequence but also to explore the local thermal field, approximated here using radial basis functions, and exploit regions of strong thermals to gain altitude. Excess altitude can then be traded for airspeed, enabling the glider to reach more waypoints within the set mission time. The problem therefore combines flight control with path planning, waypoint tracking, and energy management under environmental uncertainty, which motivates the use of a model-free learning approach to allow the agent to freely explore the design space. The waypoints are generated procedurally after the previous waypoint is reached. The distance and relative bearing to the next waypoint are sampled from uniform distributions $d_{wpt} \sim \mathcal{U}(50, 150)$ m and $\Delta \psi_{wpt} \sim \mathcal{U}(-2\pi,2\pi)$, respectively. The variation in glide distances imposes competing requirements on the aircraft: short waypoint distances favour agility and frequent turning, whereas long waypoint distances favour glide efficiency. No altitude gates are imposed at the waypoints. Each episode begins with an initial airspeed of 3.5 m/s and an initial altitude of 20 m. Following the same reward shaping rationale as in Section \ref{sec: ppf}, we define a composite reward function with multiple weighted components such that
\vspace{-10pt}

\begin{equation}
    R_t = \sum_i \lambda_ir_i.
\end{equation}

Through meticulous tuning, we have identified several reward components that prevents early training collapse as detailed in Appendix \ref{sec: Reward shaping}.

\begin{figure}[!tb]
	\centering
	\includegraphics[width=0.6\textwidth]{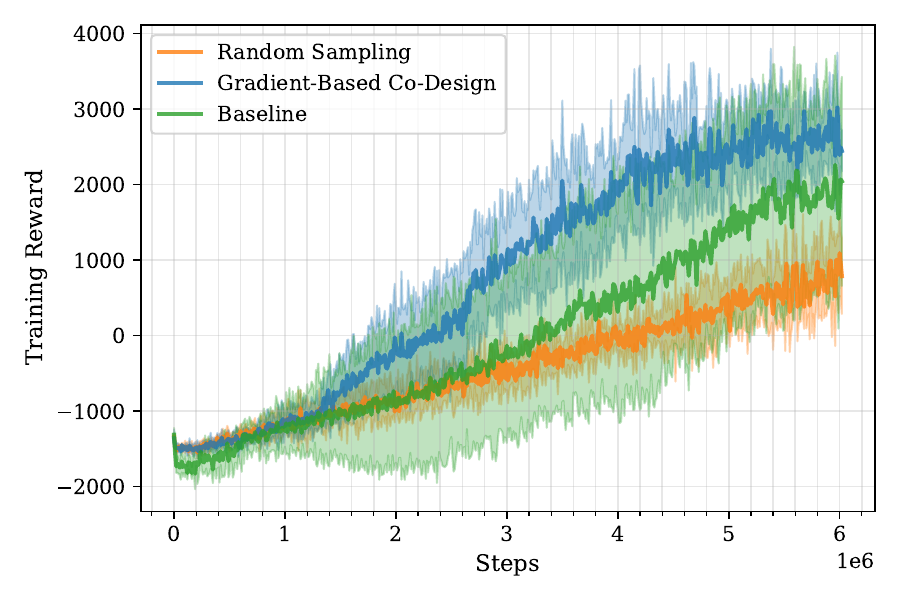}
	%\vspace{10pt}
	\caption{Reward curve of the thermal soaring glider repeated over five training attempts.}
    \vspace{20pt}
	\label{fig: co-aegis reward}
\end{figure}

\begin{figure}[!tb]
	\centering
	\includegraphics[width=0.7\textwidth]{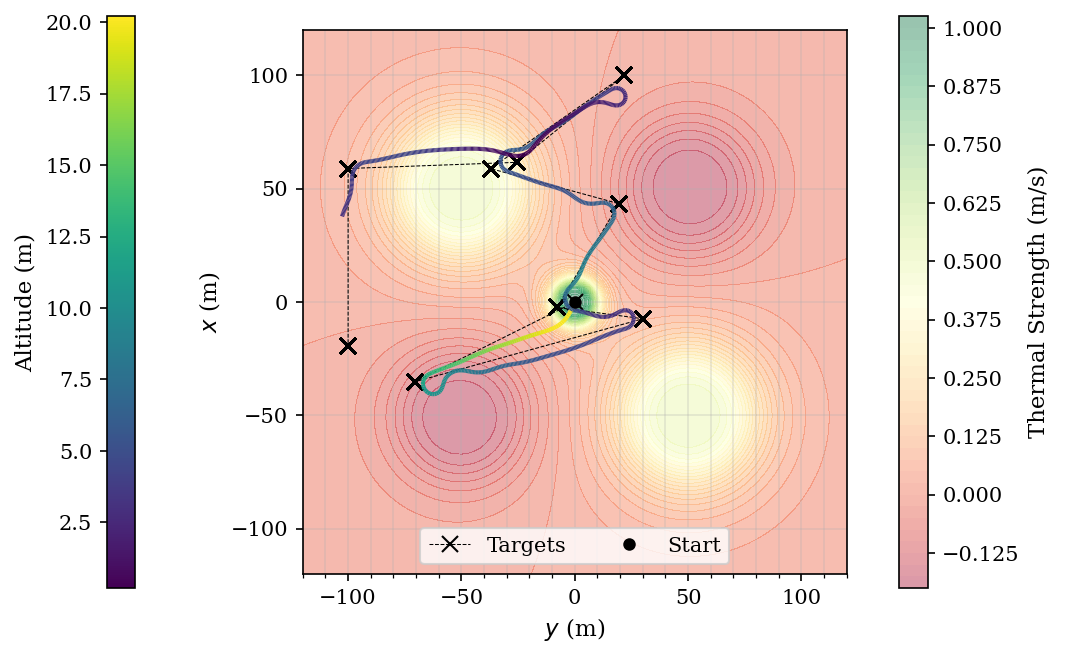}
	%\vspace{10pt}
	\caption{A sample flight trajectory using the trained policy with its optimised system parameters.}
    %\vspace{10pt}
	\label{fig: co-aegis simulation}
\end{figure}

Figure \ref{fig: co-aegis reward} shows the training reward curve over five separate training runs, benchmarked against both a policy trained on designs sampled from the design space $\mathcal{D}$ using LHS and the original baseline glider, which was optimised in open loop for minimal energy loss during flight. The co-designed glider achieves a higher reward than the benchmark glider, whose design was optimised without explicit consideration of the mission requirements. Furthermore, consistent with the results in Section \ref{sec: ppf}, guided policy training that progressively reduces the design search space is shown to be necessary for stable and effective policy learning. Without such guidance, the training process converges much slower and achieves a significantly lower reward. This is clearly evident in the sharp increase in reward of the co-design algorithm around step $2.6\times10^6$ where a design prune has occurred. Figure \ref{fig: co-aegis simulation} shows the flight trajectory of the best trained control policy. Instead of directly to its next waypoint, the agent is observed to explore its local thermal field to avoid regions of strong sinks. This training run produced candidate optimal parameters $S_\text{ref} = 0.133$ and $AR = 6.05$. Whilst the selected aspect ratio lies toward the lower end of the design space, the combination of a lower aspect ratio and a small wing area enables rapid manoeuvring through reduced roll damping and a higher optimal glide speed.

\begin{figure}[!tb]
	\centering
    \begin{minipage}{0.49\textwidth}
        \centering
    	\includegraphics[width=1\textwidth]{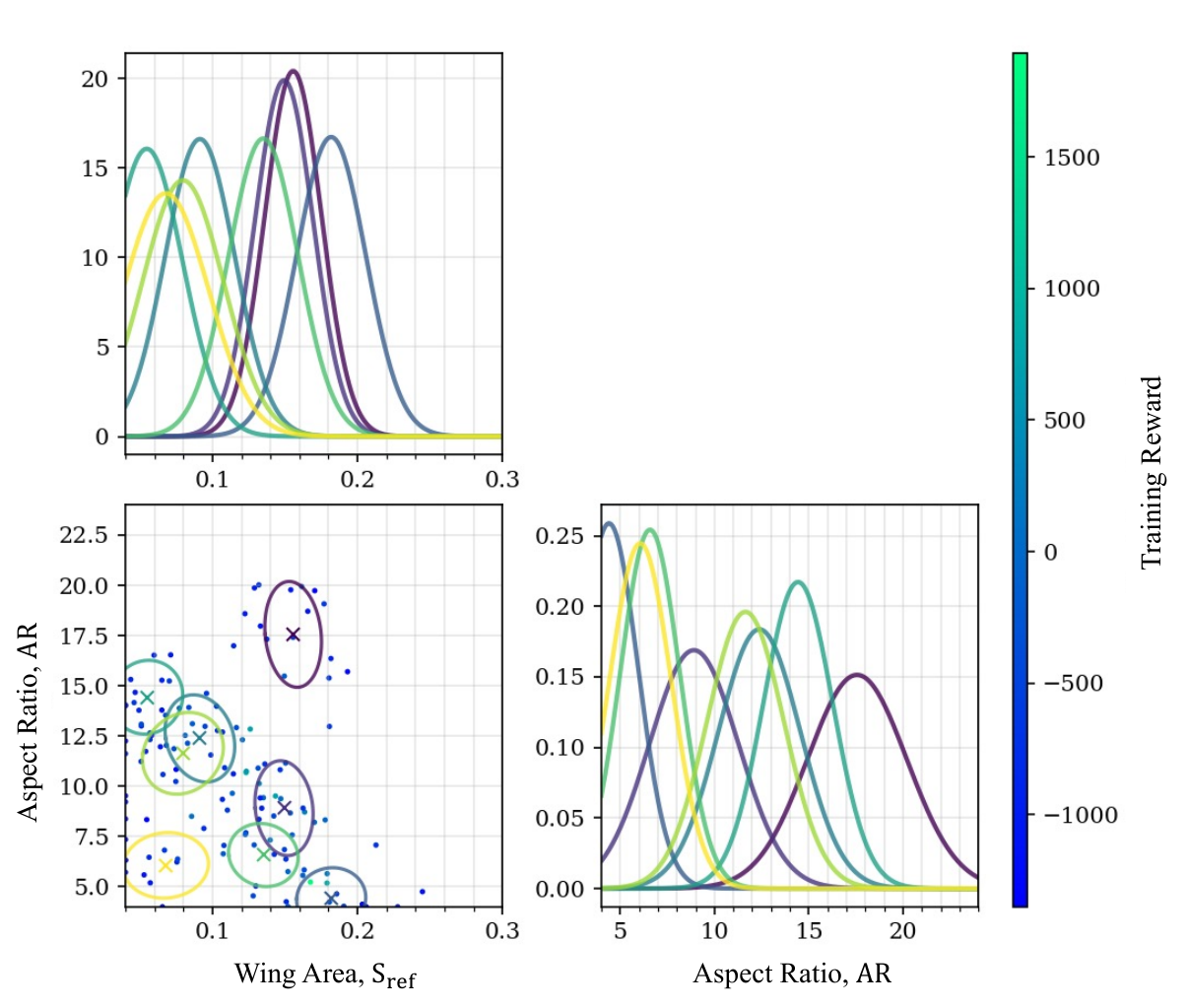}
    	%\vspace{10pt}
    	\caption{Training reward landscape at an early training stage of the glider in soaring application at step 1441792.}
    	\label{fig: co-aegis gmm early}
    \end{minipage}\hfill
    \begin{minipage}{0.49\textwidth}
        \centering
    	\includegraphics[width=1\textwidth]{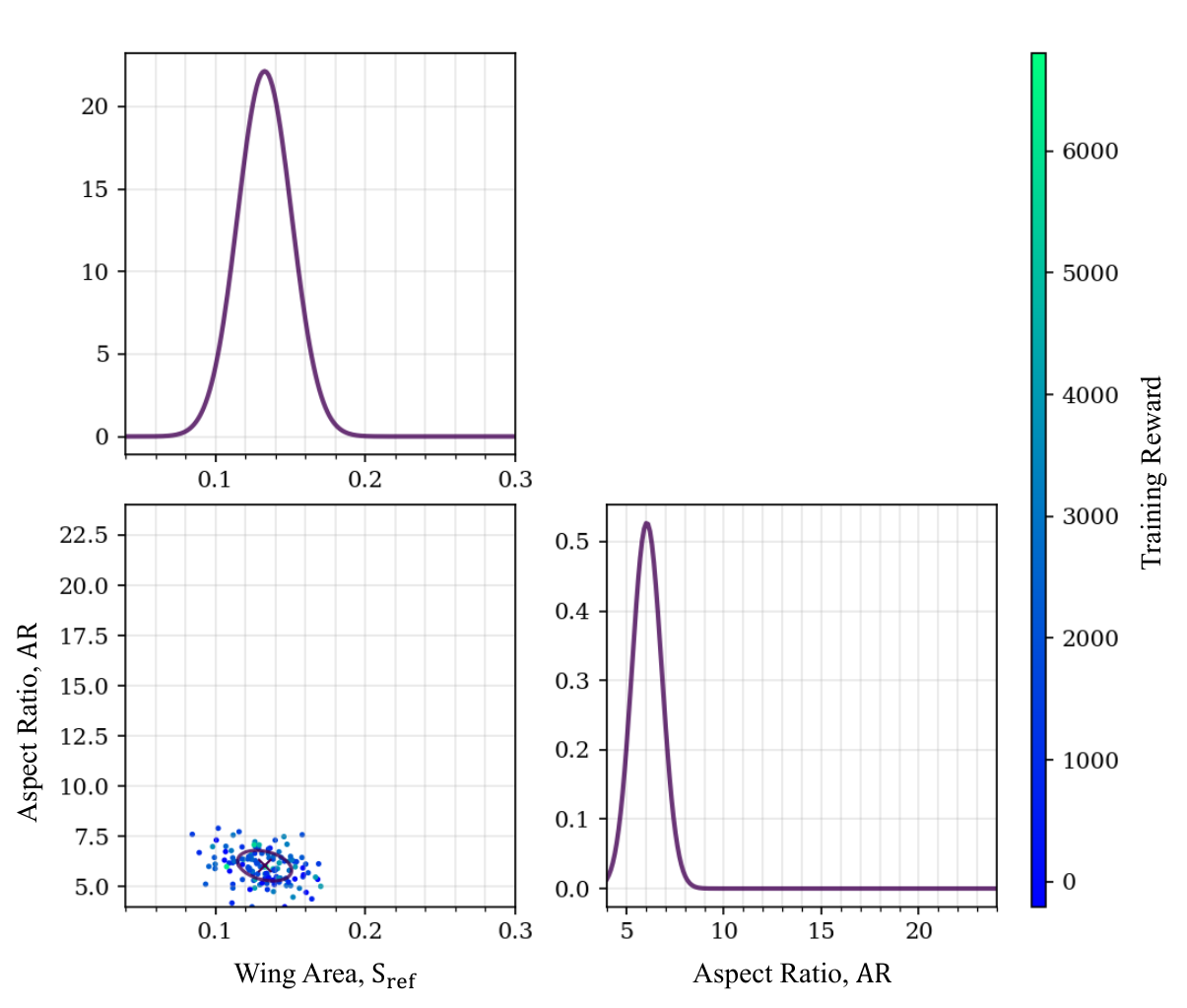}
    	%\vspace{10pt}
    	\caption{Training reward landscape before the final design distribution update of the glider in soaring application at step 6029312.}
    	\label{fig: co-aegis gmm}
    \end{minipage}
\end{figure}

Figures \ref{fig: co-aegis gmm early} and \ref{fig: co-aegis gmm} show the evolution of the design distributions during optimisation. Even at an early stage of training, the distributions shift clearly away from higher-aspect-ratio and larger wing area designs. This may be attributed both to increased roll damping and to higher induced bending loads. Larger bending loads may reduce lift efficiency if the resulting cruise condition dihedral angle becomes too high. Additionally, a smaller wing reduces the wing loading and allows the aircraft to operate at a higher optimal glide speed.

%----------------------------------------------------------------------------------------

\section{Conclusion \& Outlook}
\label{sec: conclusion and outlook}

In this study, we adapted the co-design framework \cite{Schaff_Codesign} and applied it to three case studies of increasing complexity. Beginning with a spring-mass-damper system and a pitch-plunge-flap aerofoil, the work culminated in a fixed-wing thermal-soaring mission, using a novel flexible aircraft model implemented in Mujoco. Across these case studies, the proposed model-free nested-loop co-design framework demonstrated to identify optimal joint design-control solutions in partially observed and stochastic environments. In the final fixed-wing glider case, this included not only flight control, but also waypoint tracking and path planning. The challenge and complexity of the final case study arise from both the flight model and the mission setup. The aeroelastic flight model couples the aircraft's rigid-body flight dynamics with structural degrees of freedom, while the vast design space produces substantial variation in handling qualities across candidate aircraft. At the mission level, waypoints are generated procedurally within an environment containing thermic lift and sink regions. This forces the agent to alter its flight path according to the perceived thermals in order to reach as many waypoints as possible during the set mission duration without crashing or stalling. Reward shaping played an important role in making this problem tractable by reducing training variance and encouraging desirable behaviours such as turn coordination.

More broadly, the proposed framework may be useful in a range of engineering and science applications of the kind highlighted in \cite{Alleyne_Control_Roadmap_2030}. The implementation of aeroelastic wings in Mujoco also provides an additional route for modelling flying vehicles in that environment, complementing prior work such as \cite{Vaxenburg_Fruit_fly}, and may support future development of aerial robotics applications within the Mujoco environment. More generally, the model-free co-design approach may be applicable to complex stochastic black-box optimisation problems in which system dynamics are difficult to model. One possible example is the rollout of hydrogen aircraft \cite{Muir_Sustainable_aviation_rollout}, where the economic success of the rollout depends on the interplay between the aircraft design and the acquisition and operation of the aircraft over time by relevant stakeholders. In such a setting where the system dynamics are very difficult to model, our proposed co-design inner loop could represent individual autonomous stakeholder agents that manage airline operations and fleet acquisitions, while the outer loop optimises the aircraft parameters.

Nevertheless, further investigation is needed to assess the robustness of the proposed method in highly stochastic problems. A key component of this study was reward shaping. However, in other black-box optimisation problems, the expert knowledge required to implement reward shaping or methods such as warm-start PPO \cite{Coletti_PPO_warm_start} may not be available. Future work should therefore investigate alternative approaches, including off-policy methods like SAC, which may improve sample efficiency and allow the policy optimiser to learn from all past experiences. Such methods may help stabilise training by preventing the optimiser from regressing into a local minimum. At the same time, for optimisation problems with known models, a model-free black-box approach may not be the most suitable choice,  particularly when unexpected behaviours arise that are difficult to interpret, as highlighted in Section \ref{sec: ppf}.

%----------------------------------------------------------------------------------------

\bibliography{references}
\bibliographystyle{unsrt}

%----------------------------------------------------------------------------------------

\appendix
\section{Mujoco Flight Environment}
\label{sec: Mujoco Flight Environment}

Mujoco, which stands for multi-joint dynamcis with contact, is a physics engine developed to facilitate model-based control developments \cite{Todorov_Mujoco}. Physical objects known as \textit{body} within Mujoco are defined in a kinematic tree, where each child object has a parent object. Degrees of freedom may be specified between parent and child by defining primitive \textit{joint} objects such as \textit{slide}, \textit{hinge}, \textit{ball} or \textit{free} joints. Each \textit{body} has assigned mass and inertia matrix stored within its \textit{inertial} property, along with a geometric definition in its \textit{geom} property. Whilst Mujoco offers significant capabilities in solving for nonlinear contact physics and equality constraints, the main points of interest for this work is its recursive Newton Euler algorithm implementation in computing the composite rigid body (CRB) algorithm \cite{Featherstone_CRB} for rigid multi-bodies.

This section details the implementation of a discretised wing based on \cite{Havaza_Discrete_aeroelastic} where a continuous wing was discretised into uniform span strip sections, where each section contains: elastic elements which prescribes the structural stiffness of the wing; inertial elements which defines each section's mass and inertia matrix; and aerodynamic elements which introduces aerodynamic forces that are exerted on the wing. These three elements are integrated within Mujoco by exploiting existing Mujoco physics engine features where possible and implementing new ones where necessary. As an initial implementation, several assumptions were made: the wing operates in low subsonic conditions, the wing is straight and non-tapered, any span-wise flow do not produce any aerodynamics forces nor moments, and wing-wake interactions are negligible. Post-stall behaviour is modelled using Viterna theory \cite{Tangler_Viterna} to prevent the optimiser from generating unfeasible designs or flying unphysical paths.

\begin{figure}[htb]
	\centering
	\includegraphics[width=0.6\textwidth]{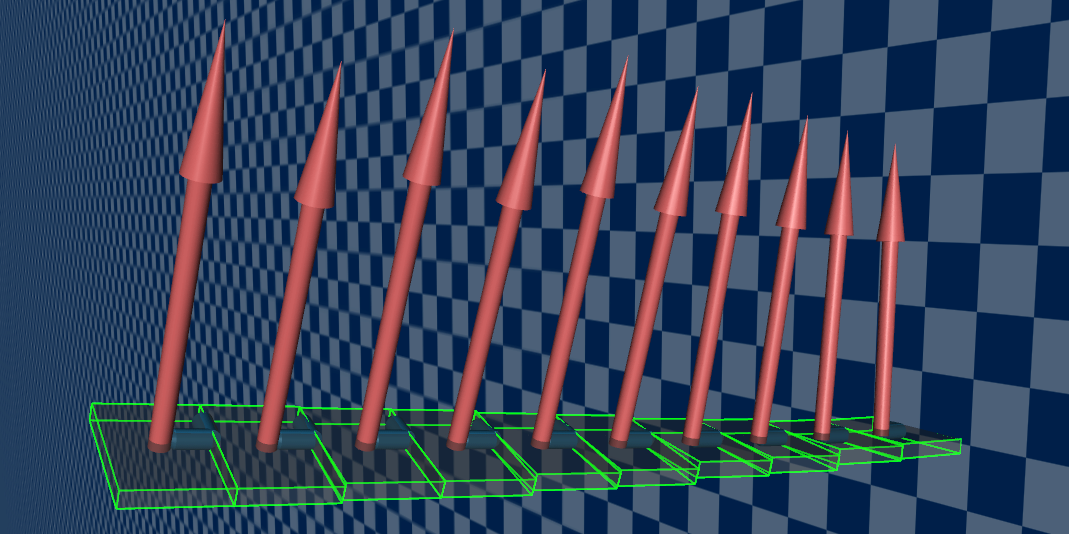}
	\vspace{10pt}
	\caption{Example discretisation of a clamped wing in Mujoco.}
	\label{fig: mujoco wing discretisation}
\end{figure}

Within the Mujoco implementation, each wing section was represented by a body which has a prescribed mass $\bar{m}$ and inertia matrix $\bar{I}$, where $\bar{\cdot}$ denotes sectional properties. Structural elasticity is modelled for out of plane bending and torsional twist using two torsion springs with linear stiffnesses, denoted as $k_{\hat{\phi}_i}$ and $k_{\hat{\theta}_i}$ respectively. In-plane bending was deemed negligible due to the significantly higher flexural stiffness in the in-plane direction of typical wing boxes. A preview of the implementation is shown in Figure \ref{fig: mujoco wing discretisation} where a straight, non-tapered wing of semi-span $b/2$ is discretised into 10 discrete strip sections each with a sectional span of $\bar{b}$. In between each section are two torsional springs denoted by blue vector joints, which provides wing out-of-plane bending $\hat{\phi}_{w_i}$ and wing torsional twist $\hat{\theta}_{w_i}$ degrees of freedom at the wing section index $i$. Aerodynamic forcing was applied using Mujoco's controller forcing callback, which applies both a 3D translational force and a 3D rotational torque at the centre mass of the object's body in the inertial frame of reference. The applied aerodynamics forces are shown using red vectors in Figure \ref{fig: mujoco wing discretisation}.

\begin{figure}[!hb]
	\centering
	\includegraphics[width=0.5\textwidth]{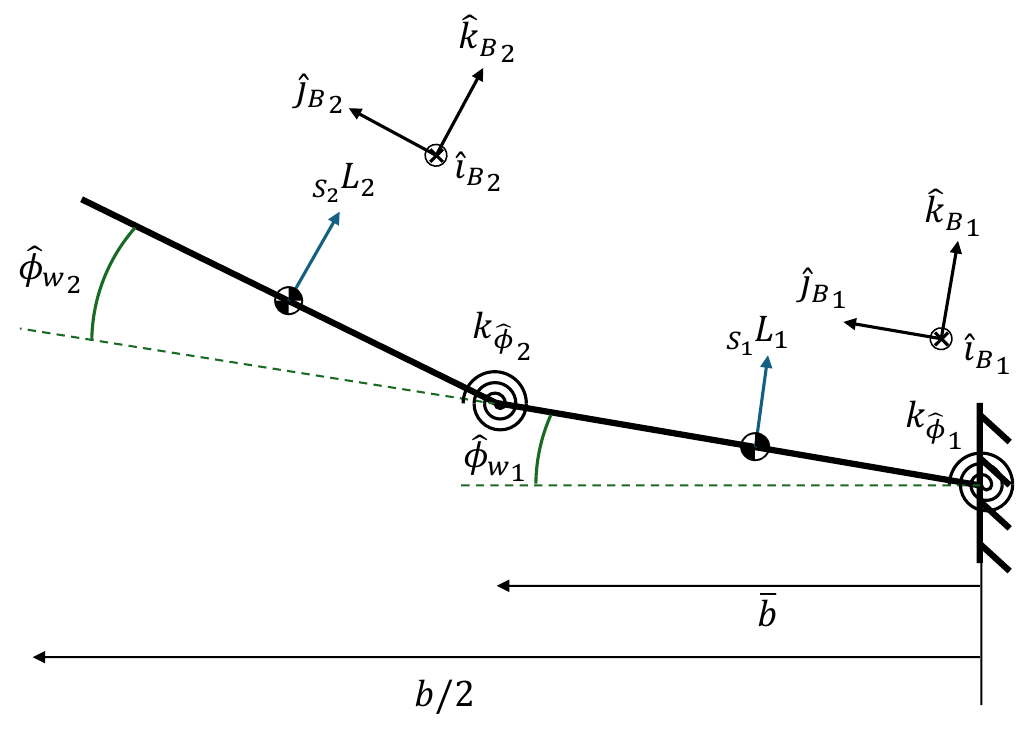}
	\vspace{10pt}
	\caption{Rear view of a 2 section discretisation of the port side wing.}
	\label{fig: wing rear view}
\end{figure}

\begin{figure}[!hb]
	\centering
	\includegraphics[width=0.5\textwidth]{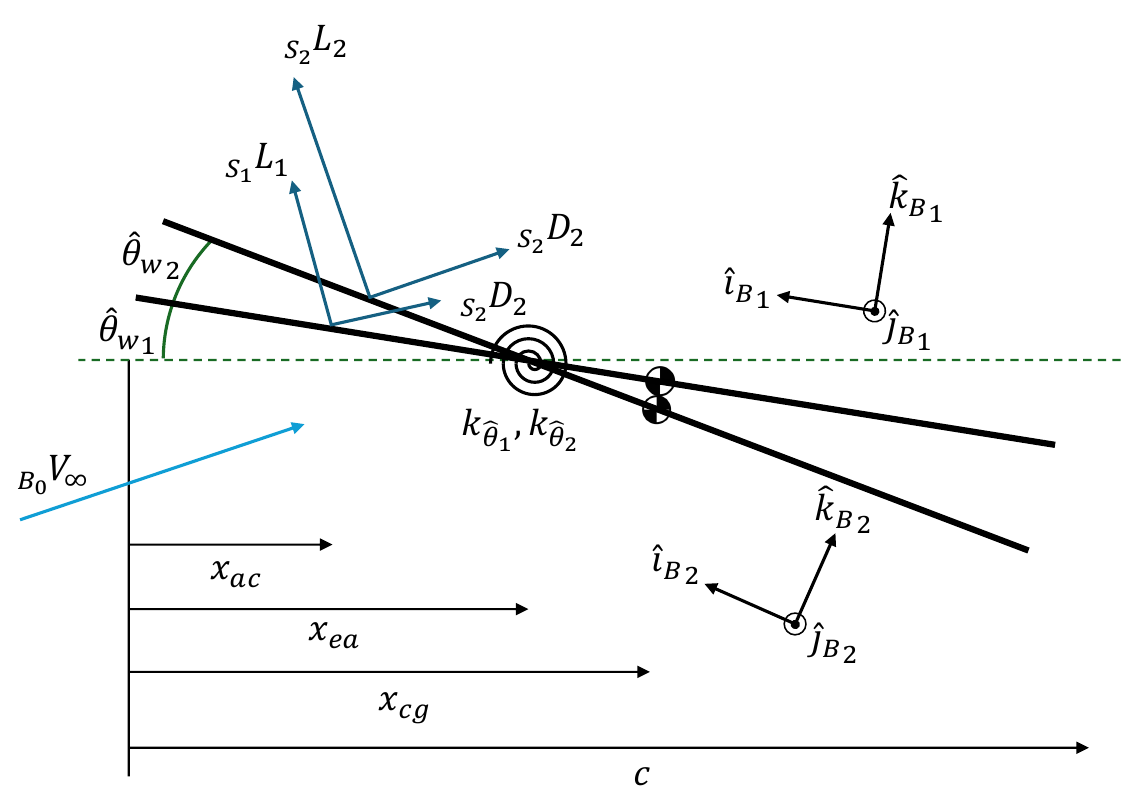}
	\vspace{10pt}
	\caption{Port side view of a 2 section discretisation.}
	\label{fig: wing side view}
\end{figure}

As shown in Figure \ref{fig: wing rear view}, the wing is discretised into uniform span sections of length $\bar{b}$ with a torsional spring of stiffness $k_{\hat{\phi}_i}$ which models out-of-plane flexural stiffness. Similar to \cite{Havaza_Discrete_aeroelastic}, a uniform lift distribution was implemented, but is transformed into an equivalent point force and torque acting at the centre of gravity that can be implemented into Mujoco. Figure \ref{fig: wing side view} shows the twist deformation of the sections and the torsional springs with stiffnesses $k_{\hat{\theta}_i}$ that model the torsional stiffness. Non-zero dihedral angle $\Lambda$ and root incident angles $i_r$ can be introduced by offsetting the neutral out-of-plane bending and twist angles by the respective angles. The longitudinal position of the spring is defined at distance $x_{ea}$ away from the leading edge and is the flexural axis of the wing. Whereas, the longitudinal position of the wing centre of mass is defined as $x_{cg}$ away from the leading edge. Due to the low subsonic flight regime and largely linear flight envelope of a glider, it is assumed that the position of the aerodynamic centre is fixed at the quarter chord point defined as $x_{ac}=c/4$. Note that the aerodynamic lift $\prescript{}{S_i}{\bar{L}_i}$ and drag $\prescript{}{S_i}{\bar{D}_i}$ forces are in the stability axes frame of wing section index $i$. The stability axes of a given wing section may not necessarily align with the oncoming freestream velocity in the body axes frame of the fuselage root $\prescript{}{B_0}{V_\infty}$, where index 0 denotes the root. Aerodynamic moments are also modelled but not shown in Figures \ref{fig: wing rear view} and \ref{fig: wing side view} for clarity. Further explanation is detailed in Section \ref{sec: Aerodynamic forcing}. The wings are then attached to a fuselage body. The fuselage is assumed to be a rigid body that does not generate any aerodynamic forcing nor moments, but has parameterisable inertial properties. The nose of the fuselage acts as the datum of the aircraft and all flight dynamics states reference that of the fuselage.

\begin{figure}[!hb]
	\centering
	\includegraphics[width=0.6\textwidth]{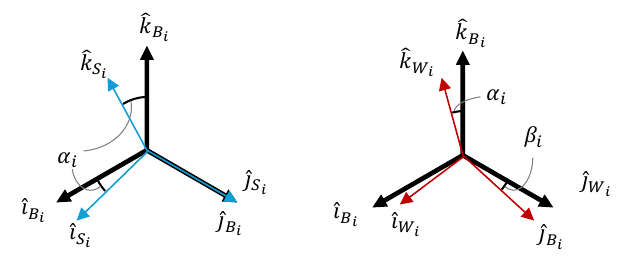}
	\vspace{10pt}
	\caption{Definition of transformation between body axes (black), stability axes (blue) and wind axes (red).}
	\label{fig: Coordinate frames}
\end{figure}

Additionally, the coordinate frame is defined as North-West-Up (NWU) to maintain consistency with Mujoco implementation requirements, with exception to Euler angles yaw $\psi_i$, pitch $\theta_i$ and roll $\phi_i$ defined using North-East-Down (NED) due to flight mechanics conventions. The transformation between the Earth-fixed \textit{worldbody} inertial frame and each body's body frame is defined with Euler angles \cite{Beard_Small_uav} where

\begin{equation}
    \left[ \begin{matrix}
        \hat{i}_{B_i} \\
        \hat{j}_{B_i} \\
        \hat{k}_{B_i}
    \end{matrix} \right]
    = R_{I1}(\psi_i)R_{12}(\theta_i)R_{2B}(\phi_i)
    \left[ \begin{matrix}
        \hat{i}_{I} \\
        \hat{j}_{I} \\
        \hat{k}_{I}
    \end{matrix} \right]
    = R_{IB}(\phi_i, \theta_i, \psi_i)
    \left[ \begin{matrix}
        \hat{i}_{I} \\
        \hat{j}_{I} \\
        \hat{k}_{I}
    \end{matrix} \right]
\end{equation}

The rotation matrices are defined as

\begin{equation}
    R_{I1}(\psi_i) =
    \left[ \begin{matrix}
        cos(\psi_i) & -sin(\psi_i) & 0 \\
        sin(\psi_i) & cos(\psi_i) & 0 \\
        0 & 0 & 1
    \end{matrix} \right]
\end{equation}

\begin{equation}
    R_{12}(\theta_i) =
    \left[ \begin{matrix}
        cos(\theta_i) & 0 & sin(\theta_i) \\
        0 & 1 & 0 \\
        -sin(\theta_i) & 0 & cos(\phi_i)
    \end{matrix} \right]
\end{equation}

\begin{equation}
    R_{2B}(\phi_i) =
    \left[ \begin{matrix}
        1 & 0 & 0 \\
        0 & cos(\phi_i) & -sin(\phi_i) \\
        0 & sin(\phi_i) & cos(\phi_i)
    \end{matrix} \right]
\end{equation}

As shown in Figure \ref{fig: Coordinate frames}, the wind axes which is aligned with the freestream velocity $\prescript{}{B_i}{\overrightarrow{U}_{\infty_i}}$ seen by the $i$ body, is two rotation transformation away from the body axes along the $z$ and $y$ axes by the sideslip angle $\beta_i$ and angle of attack $\alpha_i$ respectively \cite{Beard_Small_uav} such that

\begin{equation}
    \left[ \begin{matrix}
        \hat{i}_{W_i} \\
        \hat{j}_{W_i} \\
        \hat{k}_{W_i}
    \end{matrix} \right]
    = R_{BW}(\alpha_i, \beta_i)
    \left[ \begin{matrix}
        \hat{i}_{B_i} \\
        \hat{j}_{B_i} \\
        \hat{k}_{B_i}
    \end{matrix} \right]
\end{equation}

where the body-to-wind rotation matrix $R_{BW}(\alpha_i, \beta_i)$ is defined as

\begin{equation}
    R_{BW}(\alpha_i, \beta_i) =
    \left[ \begin{matrix}
        cos(\beta_i)cos(\alpha_i) & -sin(\beta_i)cos(\beta_i) & -sin(\alpha_i) \\
        sin(\beta_i) & cos(\beta_i) & 0 \\
        cos(\beta_i)sin(\alpha_i) & -sin(\beta_i)cos(\beta_i) & cos(\alpha_i)
    \end{matrix} \right]
\end{equation}

Whereas, the stability axes is only one rotation away from the body axes along the $y$ axis by the angle of attack $\alpha_i$ \cite{Beard_Small_uav} such that

\begin{equation}
    \left[ \begin{matrix}
        \hat{i}_{S_i} \\
        \hat{j}_{S_i} \\
        \hat{k}_{S_i}
    \end{matrix} \right]
    = R_{BS}(\alpha_i)
    \left[ \begin{matrix}
        \hat{i}_{B_i} \\
        \hat{j}_{B_i} \\
        \hat{k}_{B_i}
    \end{matrix} \right]
\end{equation}

where the body-to-stability rotation matrix $R_{BS}(\alpha_i)$ is defined as

\begin{equation}
    R_{BS}(\alpha_i) =
    \left[ \begin{matrix}
        cos(\alpha_i) & 0 & -sin(\alpha_i) \\
        0 & 1 & 0 \\
        sin(\alpha_i) & 0 & cos(\alpha_i)
    \end{matrix} \right]
\end{equation}

\subsection{Aerodynamic Forcing}
\label{sec: Aerodynamic forcing}

The aerodynamic forces applied onto body $i$ is determined by the freestream velocity seen by the body $\prescript{}{B_i}{\overrightarrow{U}_{\infty_i}}$ at its aerodynamic centre defined at the mid section span, quarter-chord point. The position of the aerodynamic centre relative to the joint is a function of the out-of-plane bending $\hat{\phi_i}$ and twist $\hat{\theta_i}$ such that

\begin{equation}
    \prescript{}{B_{i-1}}{
        \left[ \begin{matrix}
            x_i \\
            y_i \\
            z_i
        \end{matrix} \right]
    }
    = 
    \left[ \begin{matrix}
        1 & 0 & 0 \\
        0 & cos(\hat{\phi}_{w_i}) & -sin(\hat{\phi}_{w_i}) \\
        0 & sin(\hat{\phi}_{w_i}) & cos(\hat{\phi}_{w_i})
    \end{matrix} \right]
    \left[ \begin{matrix}
        cos(\hat{\theta}_{w_i}) & 0 & sin(\hat{\theta}_{w_i}) \\
        0 & 1 & 0 \\
        -sin(\hat{\theta}_{w_i}) & 0 & cos(\hat{\theta}_{w_i})
    \end{matrix} \right]
    \left[ \begin{matrix}
        x_{ea}-x_{ac} \\
        \bar{b}/2 \\
        0
    \end{matrix} \right]
\end{equation}

Its velocity relative to the joint is therefore

\begin{equation}
    \prescript{}{B_{i-1}}{
        \left[ \begin{matrix}
            u_i \\
            v_i \\
            w_i
        \end{matrix} \right]
    }
    = 
    \frac{d}{dt}\left(
        \left[ \begin{matrix}
            1 & 0 & 0 \\
            0 & cos(\hat{\phi}_{w_i}) & -sin(\hat{\phi}_{w_i}) \\
            0 & sin(\hat{\phi}_{w_i}) & cos(\hat{\phi}_{w_i})
        \end{matrix} \right]
        \left[ \begin{matrix}
            cos(\hat{\theta}_{w_i}) & 0 & sin(\hat{\theta}_{w_i}) \\
            0 & 1 & 0 \\
            -sin(\hat{\theta}_{w_i}) & 0 & cos(\hat{\theta}_{w_i})
        \end{matrix} \right]
    \right)
    \left[ \begin{matrix}
        x_{ea}-x_{ac} \\
        \bar{b}/2 \\
        0
    \end{matrix} \right]
\end{equation}

and is a function of local rotation and rotation rates. The velocities are propagated along the forward kinematic chain by Mujoco with CRB algorithm and yields the velocity of body $i$ at its aerodynamic centre relative to the fixed inertial axes in its respective body frame axes. However, the freestream velocity seen by it is only computed after accounting for the local wind velocity $\begin{bmatrix} u_W & v_W & w_W \end{bmatrix}^T$ such that

\begin{equation}
    \prescript{}{B_i}{\overrightarrow{U}_{\infty_i}} = 
    \prescript{}{B_i}{
        \left[ \begin{matrix}
            U_{\infty_i} \\
            V_{\infty_i} \\
            W_{\infty_i}
        \end{matrix} \right]
    }
    =
    \prescript{}{B_i}{
        \left[ \begin{matrix}
            u_i \\
            v_i \\
            w_i
        \end{matrix} \right]
    }
    - R_{IB}(\phi_i, \theta_i, \psi_i)^T
    \prescript{}{I}{
        \left[ \begin{matrix}
            u_W \\
            v_W \\
            w_W
        \end{matrix} \right]
    }
\end{equation}

Referencing the axes transformation shown in Figure \ref{fig: Coordinate frames}, the angle of attack and sideslip angles can be found such that

\begin{equation}
    \alpha_i = arctan \left( \frac{-\prescript{}{B_i}{W_{\infty_i}}}{\prescript{}{B_i}{U_{\infty_i}}} \right)
\end{equation}

\begin{equation}
    \beta_i = arctan \left( \frac{-\prescript{}{B_i}{V_{\infty_i}}}{\prescript{}{B_i}{U_{\infty_i}}} \right)
\end{equation}

Now transforming the velocity into the stability axes by projecting it onto the $x$-$z$ plane, the effective dynamic pressure can be defined as

\begin{equation}
    q_{\infty_i} = \frac{1}{2} \rho \left|\left| \prescript{}{B_i}{\overrightarrow{U}_{\infty_i}} \right|\right|^2cos^2(\beta_i)
\end{equation}

where $\rho$ is the air density. Therefore, the aerodynamic forces along stability axes $\hat{i}_{S_i}$, $\hat{j}_{S_i}$ and $\hat{k}_{S_i}$ are

\begin{equation} \label{eqn: wing stability forces}
    \prescript{}{S_i}{
        \left[ \begin{matrix}
            F_{\hat{i}_i} \\
            F_{\hat{j}_i} \\
            F_{\hat{k}_i}
        \end{matrix} \right]
    }
    =
    \left[ \begin{matrix}
        -\prescript{}{S_i}{D_i} \\
        0 \\
        \prescript{}{S_i}{L_i}
    \end{matrix} \right]
    =
    \left[ \begin{matrix}
        -q_{\infty_i} \bar{S}_{ref} C_{D_i}(\alpha_i, \delta_i) \\
        0 \\
        q_{\infty_i} \bar{S}_{ref} C_{L_i}(\alpha_i, \delta_i)
    \end{matrix} \right]
\end{equation}

where $\bar{S}_{ref}$ is the sectional wing area and $\delta_i$ is the trailing edge flap control surface deflection which is defined as downwards positive. Since it is assumed that spanwise flow does not generate any forces nor moments, the lateral force is assumed to be zero. The sectional lift $C_{L_i}$ and drag $C_{D_i}$ coefficients were computed using Viterna theory \cite{Tangler_Viterna} which models post-stall behaviours based on aerofoil characteristics. The drag and lift coefficients are defined as

\begin{equation}
    C_{D_i} = B_1 sin^2(\alpha_{e_i}) + B_2 cos(\alpha_{e_i})
\end{equation}
\begin{equation}
    C_{L_i} = A_1 sin(2\alpha_{e_i}) + A_2 \frac{cos^2(\alpha_{e_i})}{sin(\alpha_{e_i})}
\end{equation}

where

\begin{equation}
    C_{d_{max}} = 1.11 + 0.018AR
\end{equation}
\begin{equation}
    B_1 = C_{d_{max}}
\end{equation}
\begin{equation}
    B_2 = \frac{C_{d_{stall}} - C_{d_{max}} sin^2(\alpha_{stall})} {cos(\alpha_{stall})}
\end{equation}

\begin{equation}
    A_1 = B_1/2
\end{equation}
\begin{equation}
    A_2 = \left( C_{l_{stall}} - C_{d_{max}} sin(\alpha_{stall}) cos(\alpha_{stall}) \right)
    \frac{sin(\alpha_{stall})}{cos^2(\alpha_{stall})}
\end{equation}

\begin{equation}
    \alpha_{stall} = \frac{C_{l_{stall}} - C_{l_0}}{C_{l_\alpha}}
\end{equation}

The aerofoil drag coefficient at stall $C_{d_{stall}}$, linear lift curve slope $C_{l_\alpha}$ and its lift coefficient at zero angle of attack $C_{l_0}$ are properties of the selected aerofoil. The effective angle of attack is defined as

\begin{equation}
    \alpha_{e_i} = \alpha_i + \delta_i  \frac{C_{l_\delta}}{C_{l_\alpha}}
\end{equation}

where $C_{l_\delta}$ is the lift curve slope of its trailing edge flap. An effective angle of attack was implemented instead of the linearised summation of lift coefficient components to model near-stall behaviours of trailing edge flaps, where further positive flap deflection would exacerbate stall instead of further increasing lift generation.

Only pitching moment is considered for aerodynamic moments. The moments around the stability axes $\hat{i}_{S_i}$, $\hat{j}_{S_i}$ and $\hat{k}_{S_i}$ are

\begin{equation}
    \prescript{}{S_i}{
        \left[ \begin{matrix}
            M_{\hat{i}_i} \\
            M_{\hat{j}_i} \\
            M_{\hat{k}_i}
        \end{matrix} \right]
    }
    =
    \left[ \begin{matrix}
        0 \\
        -\prescript{}{S_i}{M_i} \\
        0
    \end{matrix} \right]
    =
    \left[ \begin{matrix}
        0 \\
        -q_{\infty_i} \bar{S}_{ref} c C_{M_i} \\
        0
    \end{matrix} \right]
\end{equation}

The moments coefficient $C_{M_i}$ is defined as

\begin{equation}
    C_{M_i} = C_{m_0} + C_{m_\delta} \delta_i
\end{equation}

where $C_{m_0}$ is the aerofoil zero-lift pitching moment coefficient and $C_{m_\delta}$ is the pitching moment slope induced by the trailing edge flaps.

Once the aerodynamic forces and moments have been obtained in the stability axes, the forces and moments are then transformed into the body axes frame such that

\begin{equation}
    \prescript{}{B_i}{
        \left[ \begin{matrix}
            F_{\hat{i}_i} \\
            F_{\hat{j}_i} \\
            F_{\hat{k}_i}
        \end{matrix} \right]
    }
    = R_{BS}(\alpha_i)
    \prescript{}{S_i}{
        \left[ \begin{matrix}
            F_{\hat{i}_i} \\
            F_{\hat{j}_i} \\
            F_{\hat{k}_i}
        \end{matrix} \right]
    }
\end{equation}

\begin{equation}
    \prescript{}{B_i}{
        \left[ \begin{matrix}
            M_{\hat{i}_i} \\
            M_{\hat{j}_i} \\
            M_{\hat{k}_i}
        \end{matrix} \right]
    }
    = R_{BS}(\alpha_i)
    \prescript{}{S_i}{
        \left[ \begin{matrix}
            M_{\hat{i}_i} \\
            M_{\hat{j}_i} \\
            M_{\hat{k}_i}
        \end{matrix} \right]
    }
\end{equation}

In order to implement aerodynamics physics as a control force callback in Mujoco, it must be offset from the aerodynamic centre to the centre of gravity of the body such that

\begin{equation}
    \prescript{}{B_{CG_i}}{
        \left[ \begin{matrix}
            F_{\hat{i}_i} \\
            F_{\hat{j}_i} \\
            F_{\hat{k}_i}
        \end{matrix} \right]
    }
    =
    \prescript{}{B_{i}}{
        \left[ \begin{matrix}
            F_{\hat{i}_i} \\
            F_{\hat{j}_i} \\
            F_{\hat{k}_i}
        \end{matrix} \right]
    }
\end{equation}

\begin{equation}
    \prescript{}{B_{CG_i}}{
        \left[ \begin{matrix}
            M_{\hat{i}_i} \\
            M_{\hat{j}_i} \\
            M_{\hat{k}_i}
        \end{matrix} \right]
    }
    =
    \prescript{}{B_i}{
        \left[ \begin{matrix}
            M_{\hat{i}_i} \\
            M_{\hat{j}_i} \\
            M_{\hat{k}_i}
        \end{matrix} \right]
    }
    +
    \left[ \begin{matrix}
        x_{cg} - x_{ac} \\
        0 \\
        0
    \end{matrix} \right]
    \times
    \prescript{}{B_i}{
        \left[ \begin{matrix}
            F_{\hat{i}_i} \\
            F_{\hat{j}_i} \\
            F_{\hat{k}_i}
        \end{matrix} \right]
    }
\end{equation}

followed by a transformation into the inertial axes

\begin{equation}
    \prescript{}{I_{CG_i}}{
        \left[ \begin{matrix}
            F_{\hat{i}_i} \\
            F_{\hat{j}_i} \\
            F_{\hat{k}_i}
        \end{matrix} \right]
    }
    = R_{IB}(\phi_i, \theta_i, \psi_i)
    \prescript{}{B_{CG_i}}{
        \left[ \begin{matrix}
            F_{\hat{i}_i} \\
            F_{\hat{j}_i} \\
            F_{\hat{k}_i}
        \end{matrix} \right]
    }
\end{equation}

\begin{equation}
    \prescript{}{I_{CG_i}}{
        \left[ \begin{matrix}
            M_{\hat{i}_i} \\
            M_{\hat{j}_i} \\
            M_{\hat{k}_i}
        \end{matrix} \right]
    }
    = R_{IB}(\phi_i, \theta_i, \psi_i)
    \prescript{}{B_{CG_i}}{
        \left[ \begin{matrix}
            M_{\hat{i}_i} \\
            M_{\hat{j}_i} \\
            M_{\hat{k}_i}
        \end{matrix} \right]
    }
\end{equation}

which are then applied onto the wing section bodies via Mujoco's callback functionality.

\subsection{Equivalent Structural Stiffness}
\label{sec: Structural formulation}

Utilising Mujoco's built in physics features, a linear stiffness value may be specified for each degree of freedom. In order to find an equivalent spring stiffness to model a continuous beam using discretised beam sections, the end tip deflection and twist are enforced to be the same as $\hat{\phi}_{w_i}$ and $\hat{\theta}_{w_i}$ under a static aerodynamic load at the mid-span. This is to ensure convergence of wing bending and twist profile under load as the wing discretisation level increases, which is critical due to its influence on the aerodynamic forces exerted onto individual discretised sections.

\begin{figure}[!hb]
	\centering
	\includegraphics[width=0.5\textwidth]{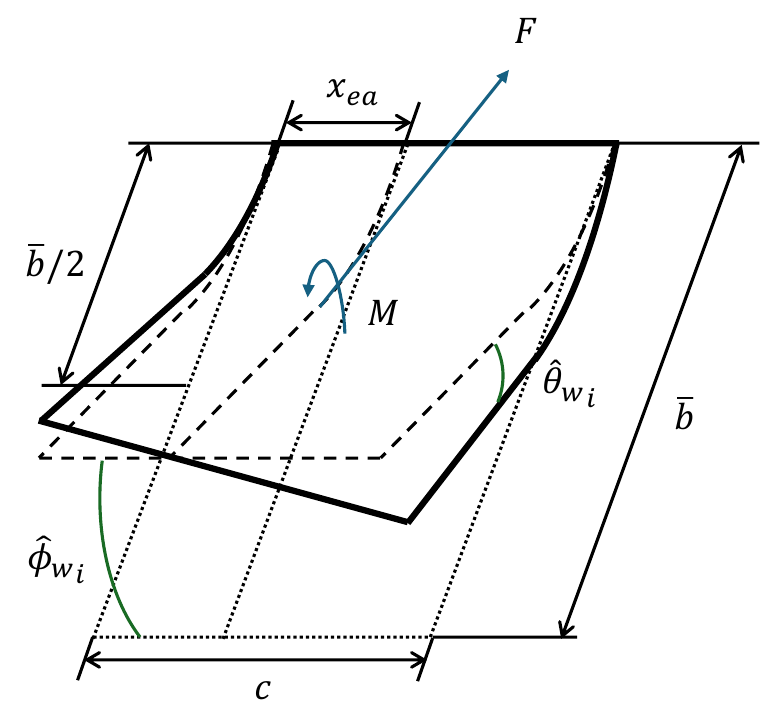}
	\vspace{10pt}
	\caption{Deflection of a continuous beam under a force and moment load at its midspan.}
	\label{fig: Equivalent stiffness beam}
\end{figure}

As shown in Figure \ref{fig: Equivalent stiffness beam}, a continuous beam with flexural stiffness of $EI_x$ and torsional stiffness of $GJ$ under a force and moment applied to its mid-span. If the tip bending deflection and twist are enforced as $\hat{\phi}_{w_i}$ and $\hat{\theta}_{w_i}$ respectively, it would yield

\begin{equation}
    \hat{\phi}_{w_i} = \frac{F\bar{b}^2}{8EI_x}
\end{equation}
\begin{equation}
    \hat{\theta}_{w_i} = \frac{M\bar{b}}{2GJ}
\end{equation}

Therefore, this would yield an equivalent linear spring stiffness in a discretised beam section as

\begin{equation}
    k_{\hat{\theta}_i} = \frac{4EI_{x}}{\bar{b}}
\end{equation}
\begin{equation}
    k_{\hat{\phi}_i} = \frac{2GJ}{\bar{b}}
\end{equation}

The area moment $I_x$ and polar of inertia $J$ was defined by approximating the wing as a homogenous rectangular thin beam element such that

\begin{equation} \label{eqn: wing area moment of inertia}
    I_x = \frac{c\bar{h}^3}{12}
\end{equation}
\begin{equation} \label{eqn: wing polar moment of inertia}
    J = \frac{c\bar{h}(c^2+\bar{h}^2)}{12}
\end{equation}

where $\bar{h}$ is the mean aerofoil height across its chord $c$. The equivalent stiffnesses are then implemented as joint stiffnesses on the joints connecting the wing sections.

\subsection{Generalised Inertial Forces}
\label{sec: Inertial formulation}

To ensure tractable symbolic calculations, \cite{Havaza_Discrete_aeroelastic} implemented a generalised cuboid inertial element. Similarly, to improve computational efficiency during preliminary studies, the inertial element of a thin wing section is generalised as a cuboid element with a longitudinal offset to allow for parameterisable centre of gravity location. Its inertia matrix is defined as

\begin{equation} \label{eqn: wing inertia}
    \bar{I} = \frac{\bar{m}}{12}
    \left[ \begin{matrix}
        \bar{b}^2 + \bar{h}^2 & 0 & 0 \\
        0 & \bar{h}^2 + c^2 & 0 \\
        0 & 0 & \bar{b}^2 + c^2
    \end{matrix} \right]
\end{equation}

Its sectional mass assumes the cuboid element is made of homogeneous material such that

\begin{equation} \label{eqn: wing mass}
    \bar{m} = \rho_w \bar{b}\bar{h}c
\end{equation}

where $\rho_w$ is the density of the wing. The inertia matrix, mass and position of the cuboid inertial element is implemented in Mujoco via its \textit{inertial} property. The inertial element is always defined with the local body axes orientation.

\section{Reward Shaping for Thermal Soaring Mission}
\label{sec: Reward shaping}

Due to the complexity of the environment, a sparse waypoint-based reward may result in poor training or even training collapse if the agent never achieves stable flight in the first place. We have introduced the following reward components to help guide the training process:

\begin{itemize}
    \item End reward $r_{end}$ - The agent receives a reward of 300 for every waypoint it reaches but is penalised by 1200 if it crashes or stalls. Such a balance was struck such that the agent is motivated to cruise at a higher speed to reach as many waypoints as possible within the time limit but not sacrifice all its altitude in the process.
    \item Angle of attack penalty $r_{\alpha}$ - The agent is penalised in proportion to the negative square its angle of attack $\alpha^2$ to prevent stalls. It should be noted that stall may occur during optimal flights around thermals and should not be excessively penalised.
    \item Pitch penalty $r_\theta$ - The agent receives a penalty proportion to the square of its pitch $\theta^2$. This is to penalise extreme attitudes but not to limit an agent's ability to climb or descend rapidly in response to the environment.
    \item Sideslip penalty $r_\beta$ - The agent receives a penalty equal to the absolute value of its sideslip angle $|\beta|$ in radians. Maintaining zero sideslip is key in reducing unnecessary drag on the aircraft and to ensure coordinated turns.
    \item Heading penalty $r_\psi$ - The agent receives a penalty equal to the absolute value of its relative bearing to waypoint $|\Delta \psi_{wpt}|$ to prompt the agent towards its objective.
    \item Pitch rate penalty $r_q$ - The agent is penalised for the absolute value of its pitch rate $|q|$ in radians. It was observed that due to the flexible aircraft model, the agent is prone to excite oscillatory behaviours in a manner similar to pilot induced oscillations.
    \item Thermal reward $r_{w_W}$ - The agent receives a reward equal to the value of the local thermal strength $\prescript{}{I}{w_W}$ to motivate the agent to stay within thermals and to avoid sinks.
    \item Sink rate penalty $r_{\dot h}$ - The agent penalty proportional to its sink rate $-\dot{h}$ to discourage excessive sacrifice of glide ratio for airspeed.
    \item Waypoint closure rate $r_{\Delta d}$ - The agent is penalised by the difference in distance to the next waypoint between the last time step and the current time step $d_{t-1} - d_t$. This encourages the agent to fly towards its waypoint but $r_\psi$, it does not penalise for loitering over thermals.
\end{itemize}

The corresponding reward weights are shown in Table \ref{tab: Reward weights}.

\begin{table}[!ht]
	\centering
	\caption{Weights of the different applied reward components.}
	\vspace{10pt}
	\begin{tabular}{cc}
    \hline
    Weight               & Value          \\ \hline
    $\lambda_{end}$      &  1.00          \\
    $\lambda_\alpha$     &  1.00          \\
    $\lambda_\theta$     &  1.00          \\
    $\lambda_\beta$      &  2.00          \\
    $\lambda_\psi$       &  0.10          \\
    $\lambda_q$          &  0.01          \\
    $\lambda_{w_W}$      &  1.00          \\
    $\lambda_{\dot{h}}$  &  2.00          \\
    $\lambda_{\Delta d}$ & 10.00          \\  \hline
    \end{tabular}
    \label{tab: Reward weights}
\end{table}

\section{Co-Design Hyperparameters}
\label{sec: codesign hyperparameters}

\begin{table}[!ht]
	\centering
	\caption{List of co-design hyperparameters used for the three case studies.}
	\vspace{10pt}
	\begin{tabular}{cccc}
    \hline
    Hyperparameter           & Spring-Mass-Damper & Pitch-Plunge-Flap & Thermal Soaring          \\ \hline
    $N_\text{total steps}$   &  300000  &  600000  &  6000000  \\
    $N_\text{env}$           &  16      &  16      &  128      \\
    $N_\text{design}$        &  8       &  8       &  16       \\
    $N_\text{freeze}$        &  10      &  10      &  10       \\
    $N_\text{inner}$         &  3       &  3       &  2        \\
    $N_\text{prune}$         &  5       &  10      &  10       \\
    $\alpha_\text{prune}$    &  0.5     &  0.5     &  0.5      \\
    $N_\text{step}$          &  256     &  256     &  512      \\ \hline
    \end{tabular}
    \label{tab: codesign hyperparameters}
\end{table}

\end{document}

%% file: preamble.tex
\usepackage[utf8]{inputenc}
\usepackage[T1]{fontenc}

\usepackage{graphicx}
\usepackage[top=2.54cm,bottom=2.54cm, left=2.54cm, right=2.54cm]{geometry}
\usepackage{fancyhdr}
\usepackage{hyperref}
\usepackage{lastpage}
\usepackage{multicol}
\usepackage{titling}
\usepackage{cite}
\usepackage{amsmath}
\usepackage{amsfonts}
\usepackage{amssymb}
\usepackage{verbatim}
\usepackage{floatpag}
\usepackage{siunitx}
\usepackage{amsmath,lipsum}
\usepackage{multicol}
\usepackage{booktabs}
\usepackage{float}
\usepackage{pgfplots}
\usepackage{bm}
\usepackage{svg}
\usepackage{nameref}
\usepackage{longtable}
\usepackage{pdfpages}
\usepackage{subcaption}
\usepackage{tabularx}
\usepackage{multirow}
\usepackage{graphicx}
\usepackage{caption}
\usepackage{subcaption}
\usepackage{adjustbox}

\usepackage{graphicx}
\usepackage{caption}

\usepackage{lscape}
\usepackage{rotating}
\usepackage{tabularx}

\usepackage{parskip}
\usepackage{gensymb}
\usepackage{epsfig}
\usepackage{makecell}
\usepackage{enumitem}

\usepackage{mathtools}
\usepackage{amsmath}
\usepackage{algorithm}
\usepackage{algpseudocode}
\usepackage{subcaption}
\usepackage{placeins}

\usepackage{authblk}

\newcommand{\varline}{\leavevmode\leaders\hrule height 3.2pt depth -2.8pt\hfill\kern0pt}